\documentclass[usenatbib]{mnras}

\usepackage{natbib}
\usepackage{delarray}
\usepackage{color}
\usepackage{amsmath}
\usepackage{amssymb}
\usepackage{here}
\usepackage{graphicx}
\usepackage[utf8]{inputenc}
\usepackage{float}
\usepackage{epsfig}
\usepackage{tensor}
\usepackage{fixltx2e} 
\usepackage{xspace}

\usepackage[T1]{fontenc}
\usepackage{ae,aecompl}

\usepackage{enumitem} 

\newcommand{\mach}{\mathcal{M}}
\newcommand{\xvect}{\mathbf{x}}
\newcommand{\kvect}{\mathbf{k}}
\newcommand{\be}{\begin{equation}} \newcommand{\ee}{\end{equation}}

\newcommand{\solarmass}{\mathrm{M}_{\sun}}
\newcommand{\lang}{\left\langle} \newcommand{\rang}{\right\rangle}
\newcommand{\pc}{\mathrm{pc}}
\newcommand{\lbra}{\left(}
\newcommand{\rbra}{\right)}
\newcommand{\dderiv}{\mathrm{d}}

\newcommand{\acknowledgments}{\begin{small}\section*{Acknowledgments}\end{small}}
\newcommand\altaffilmark[1]{$^{#1}$}
\newcommand\altaffiltext[1]{$^{#1}$}

\usepackage{tikz}
\usetikzlibrary{shapes.geometric, arrows, calc}
\tikzstyle{startstop} = [rectangle, rounded corners, minimum width=3cm, minimum height=1cm,text width=4cm,text centered, draw=black, fill=red!30]
\tikzstyle{io} = [trapezium, trapezium left angle=70, trapezium right angle=110, minimum width=4cm, minimum height=1cm,text width=3cm, text centered, draw=black, fill=blue!30]
\tikzstyle{process} = [rectangle, minimum width=3cm, minimum height=1cm, text width=5cm,text centered, draw=black, fill=orange!30]
\tikzstyle{decision} = [diamond, minimum width=3cm, text width=2cm, minimum height=1cm, text centered, draw=black, fill=green!30]
\tikzstyle{arrow} = [thick,->,>=stealth]

\newcommand{\genturb}{Paper II\xspace}
\newcommand{\CMFIMF}{Paper I\xspace}
\newcommand{\myquote}[1]{``#1''}

\title{Star Formation in a Turbulent Framework: From Giant Molecular Clouds to Protostars}

\author[Guszejnov \&\ Hopkins]{
\parbox[t]{\textwidth}{ D\'avid Guszejnov\altaffilmark{1}\thanks{E-mail:guszejnov@caltech.edu} and Philip F. Hopkins\altaffilmark{1}}
\vspace*{6pt} \\
\altaffiltext{1}{TAPIR, Mailcode 350-17, California Institute of Technology, Pasadena, CA 91125, USA} \\
}

\date{To be submitted to MNRAS, \today \vspace{-0.6cm}}
\begin{document}
\maketitle
\label{firstpage}

\begin{abstract}

Turbulence is thought to be a primary driving force behind the early stages of star formation. In this framework large, self gravitating, turbulent clouds fragment into smaller clouds which in turn fragment into even smaller ones. At the end of this cascade we find the clouds which collapse into protostars. Following this process is extremely challenging numerically due to the large dynamical range, so in this paper we propose a semi analytic framework which is able to model star formation from the largest, giant molecular cloud (GMC) scale, to the final protostellar size scale. Due to the simplicity of the framework it is ideal for theoretical experimentation to explore the principal processes behind different aspects of star formation, at the cost of introducing strong assumptions about the collapse process. The basic version of the model discussed in this paper only contains turbulence, gravity and crude assumptions about feedback, nevertheless it can reproduce the observed core mass function (CMF) and provide the protostellar system mass function (PSMF), which shows a striking resemblance to the observed IMF, if a non-negligible fraction of gravitational energy goes into turbulence. Furthermore we find that to produce a universal IMF protostellar feedback must be taken into account otherwise the PSMF peak shows a strong dependence on the background temperature.

\end{abstract}

\begin{keywords}
stars: formation -- turbulence -- galaxies: evolution -- galaxies: star formation -- cosmology: theory
\vspace{-1.0cm}
\end{keywords}

\section{Introduction}\label{sec:intro}

	Finding a comprehensive description of star formation has been one of the principal challenges of astrophysics for decades. Such a model would prove invaluable to understanding the evolution of galactic structures, binary star systems and even the formation of planets.
	
\begin{figure*}
\begin {center}
\includegraphics[width=\linewidth]{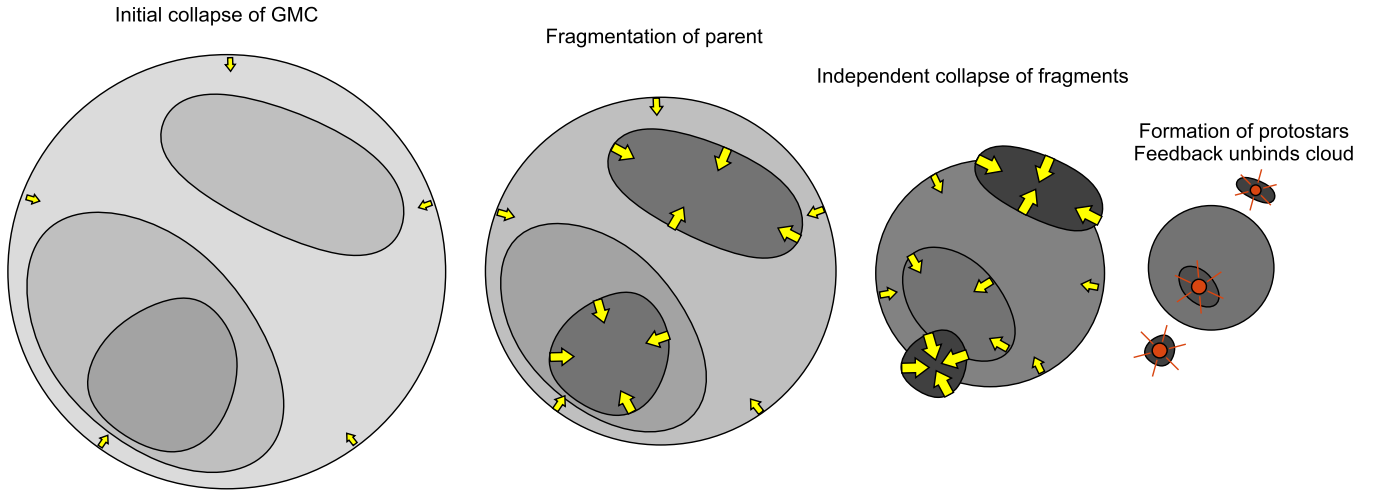}
\caption{Evolution of collapsing clouds, with time increasing from left to right (darker subregions are higher-density, arrows denote regions which are independently self-gravitating and become thicker with increasing collapse rate). As the initial cloud collapses, density fluctuations increase (because gravitational energy pumps turbulence), creating self-gravitating subregions. These then collapse independently from the parent cloud, forming protostars at the end. These protostars can provide a sufficiently strong feedback that the rest of the cloud becomes unbound and ceases to collapse.}\label{fig:fragment_fig}
\end {center}
\end{figure*}

	It has been long established that stars form from collapsed dense molecular clouds (\citealt{McKee_star_formation}). Currently the most promising candidate for a driving process is turbulence, as it can create subregions with sufficiently high density so that they become self gravitating on their own, while also exhibiting close to scale free behavior (in accordance with the observations of \citealt{Larson_law, Bolatto_2008}). These fragments are inherently denser than their parents so they collapse faster, quasi independent from their surroundings. However, once they turn into stars they start heating up the surrounding gas (by radiation, solar winds or supernova explosions) preventing it from collapsing and forming stars (see  Fig. \ref{fig:fragment_fig}). This process is inherently hierarchical so it should be possible to derive a model that follows it from the scale of the largest self gravitating clouds, the GMCs ($\sim 100\,\pc$), to the scale of protostars ($\sim 10^{-5}\,\pc$). This is not possible in direct hydrodynamic simulations due to resolution limits, but can be treated approximately in analytic and semi-analytic models.
	
	This paradigm has been explored by \cite{Padoan_theory} and \cite{Padoan_Nordlund_2002_IMF}, then made more rigorous by \cite{HC08} who attempted to create an analytic model analogous to \cite{PressSchechter}, which approximates the background density field as a Gaussian random field. A similar model was developed by \cite{Vazquez_SF_model_2012}, however that did not rely on turbulence.  Later \cite{excursion_set_ism} expanded on these works by adopting the excursion set formalism to find the distribution of the largest self gravitating structures, which was found to be very similar to the observed distribution of GMCs. Similarily \cite{core_IMF} found that the distribution of the smallest self gravitating structures fit well the observed CMF. Building on these results \cite{general_turbulent_fragment} generalized the formalism to be applicable to systems with different equations of state and turbulent properties. 
	
	Observed cores are sub-sonic and show no clear sign of fragmentation and the CMF looks very similar to the IMF apart from a factor of $\sim 3$ shift in the mass scale (\citealt{IMF_universality}). However, if no other physics is assumed other than isothermal turbulence and gravity, during the collapse the cores develop strong turbulence and eventually sub-fragment into smaller objects (\citealt{Goodwin04a, Walch12a}, for discussion see \citealt{SF_big_problems}). This implies that some additional physics must play a role, but there is no clear consensus on what it could be: magnetic fields (\citealt{Nakano_magnetic_fields, McKee_SF_theory}), radiation (\citealt{Krumholz_stellar_mass_origin}), cooling physics (\citealt{Jappsen_EQS_ref}) etc. Using a cooling physics motivated \myquote{stiff} EOS \cite{GuszejnovIMF} incorporated the time dependent collapse of the cores into the excursion set formalism and found that the distribution of protostars closely reproduced the observed IMF.
	
	These excursion set models did successfully reproduce the CMF, IMF and the GMC mass function, however they had several shortcomings. First, they did not account for the differences in formation and collapse times of clouds of different sizes (e.g. small clouds form faster and collapse faster). Secondly, the excursion set formalism describes the density field around a random Lagrangian point. This means that the spatial structure of a cloud can not be modeled directly (e.g. there is no way to find if a cloud forms binary stars). Finally, there is no self consistent excursion set model that follows from the GMC to the protostar scale (i.e. \citealt{core_IMF} covered scales between the galactic disk and cores, \citealt{GuszejnovIMF} between cores and protostars). We believe these shortcomings can be overcome by moving away from the analytic excursion set formalism and instead adopting a simple semi-analytical approach with the same random field assumption. This framework would allow us to follow the evolution self gravitating clouds while resolving both the GMC and protostellar scales and preserving spatial information. In this paper we will outline a possible candidate for such a model. 
	
	The paper is organized as follows. Sec. \ref{sec:methodology} provides a general overview of the model, including the primary assumptions and approximations and briefly outlines its numerical realization. Sec. \ref{sec:imf_cmf_evol} shows the simulated time evolution of the CMF and the protostellar system mass function (PSMF) which shows a striking similarity to the IMF. Sec. \ref{sec:IMF} also discusses the effects of having a temperature independent equation of state on the peak of the PSMF and the universality of the IMF. Finally, Sec. \ref{sec:conclusions} discusses the results and further applicability of the model.

\section{Methodology}\label{sec:methodology}
	
	In short, instead of doing a detailed hydrodynamical simulation involving gravity and radiation, our model assumes a simple stationary model for the density field, collapse of structures at constant virial parameter and an equation of state that depends on cloud properties. Starting from a GMC sized cloud it evolves the density field as the cloud collapses and pumps turbulence (this is not a bad approxiation, see \citealt{Brant_turb_pumping, Murray_star_formation, Murray_2015_turb_sim}). Note that our assumptions do not necessarily mean that all clouds have supersonic turbulence. \genturb has shown that if a medium has a \myquote{stiff} equation of state ($\gamma>4/3$), then turbulence is dampened during collapse. Since it is observed that dense, low mass cores are subsonic while high mass, low density clouds are supersonic some form of physics is needed to remove the turbulent energy. For that purpose we are using an equation of state that becomes stiff at high densities, which in combination with the constant virial parameter assumption makes dense clouds sub-sonic, arresting fragmentation.
	
	In the model, at each time step we search for self gravitating structures which we treat as new fragments, for which the process is repeated in recursion until a substructure is found that collapses to protostellar scale without fragmenting. Our assumptions will be discussed in more detail in the following subsections while a step-by-step description of the algorithm is provided in Appendix \ref{sec:algorithm}.
	
	Our model is a modified version of the excursion set model used by \cite{GuszejnovIMF} (henceforth referred to as \CMFIMF) using the theoretical foundation of \cite{general_turbulent_fragment} (henceforth referred to as \genturb). Due to the significant overlap between models we show only the essential equations and emphasize the differences and their consequences. If the reader is familiar with \CMFIMF we suggest skipping to Sec. \ref{sec:model_diff}.

\subsection{The Density Field}

\begin{figure}
\begin {center}
\includegraphics[width=\linewidth]{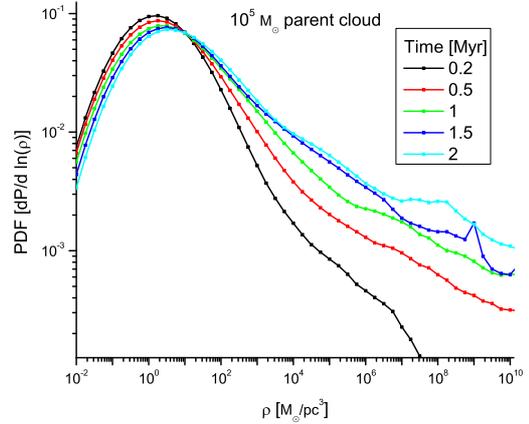}
\caption{Time evolution of the distribution of density in a parent GMC of $10^5\,\solarmass$. This is a mass weighted average of the density distribution of all substructures in the parent cloud (which are all assumed to be lognormal with different parameters), thus the low mass end is set by the lowest density structure which is the parent cloud while the high mass end is a power law due to the power law like distribution of fragments (see Fig. \ref{fig:bound_struct_evol}). There is also a clear trend as the high mass end tail rises in time. This is caused by the formation of new self gravitating substructures (\protect\citealt{Federrath_density_distrib}).}\label{fig:dens_distrib}
\end {center}
\end{figure}

It is known that the density field in the cases of both sub and supersonic, isothermal flows follows approximately lognormal statistics (for corrections see \cite{Hopkins_isothermal_turb}). This means that if we introduce the density contrast $\delta(\xvect)=\ln\left[\rho(\xvect)/\rho_0\right]+S/2$, with $\rho(\xvect)$ as the local density, $\rho_0$ as the mean density and $S$ as the variance of $\ln\rho$, it would follow a close to Gaussian distribution\footnote{It is a common misconception that analytical models such as the one presented in this paper take the total density distribution to be purely lognormal. While the density distribution in each cloud/fragment is indeed assumed to be locally lognormal on a single timestep, these have different means and deviations (see Eq. \ref{eq:S_def}) depending on their initial conditions and time, which means that the total distribution will be different. If we measure the density distribution in our calculations (see Fig. \ref{fig:dens_distrib}), we find it is approximately lognormal at low densities (set by the lowest density structure: the parent cloud), while the high mass end becomes a power law as it is a mass weighted average of the distributions for different substructures whose mass distribution is a power law (see Fig. \ref{fig:bound_struct_evol}).}, thus
\be
P(\delta|S)\approx\frac{1}{2\pi S}\exp{\left(-\frac{\delta^2}{2S}\right)}.
\ee
It is a property of normal and lognormal random variables that a linear functional of these variables will also be normal/lognormal, thus the averaged density in a region has lognormal equilibrium statistics whose properties are prescribed by turbulence. Following \genturb this yields
\be
S(\lambda)=\int_{0}^{\lambda}{\Delta S(\lambda)d\ln{\lambda}}\approx\int_{0}^{\lambda}{\ln{\left[1+b^2\mach^2\lbra \lambda\rbra\right]}\dderiv\ln{\lambda}},
\label{eq:S_def}
\ee
where $\lambda$ is the averaging scale, $\mach\left(\lambda\right)$ is the Mach number of the turbulent velocity dispersion on scale $\lambda$ and $b$ is the fraction of the turbulent kinetic energy in compressive motions, which we take to be about $1/2$ (this is appropriate for an equilibrium mix of driving modes, see \citealt{Federrath_turbulence_compressive_PDF} for details. \CMFIMF experimented with $b\sim 1/4-1$ and found no qualitative differences).

It is important to note that although $\rho$ is lognormal which means $\delta$ is Gaussian, there is significant spatial correlation (i.e. $\rho$ can not change instantly over arbitrarily small spatial intervals) so it is not possible to model the density field as a spatially independent random field. To circumvent this issue we solve the problem in Fourier space since $\delta\left(k\right)$ is also lognormal, while there is little correlation between modes so it is acceptable to assume them to be independent (note: having correlated modes in Fourier space introduces only mild effects on the final mass functions, see Appendix A of \genturb for details). Combined with the fact that the number of modes in the $[k,k+\dderiv k]$ range is $\dderiv N(k)=\lbra 4\pi k^2\dderiv k\rbra n_k$, where $n_k$ is the mode density, we get the variance for an individual density contrast mode is
\be
S_{\rm mode}(\kvect)=\frac{\ln(1+b^2\mach(k)^2)}{ 4 \pi k^3 n_k}.
\label{eq:S_per_mode}
\ee

\genturb showed that to realize a steady state density contrast field with such variance and zero mean, the Fourier component $\delta(\kvect,t)$ must evolve as
\be
\label{eq:delta_evol}
\delta(\kvect,t+\Delta t)=\delta(\kvect,t)\left(1-\Delta t/\tau_{k}\right)+\mathcal{R}\sqrt{2 S_{\rm mode}(\kvect)\Delta t/\tau_{k}},
\ee
where $\mathcal{R}$ is a Gaussian random number with zero mean and unit variance while $\tau_{k}\sim v_t(k)/\lambda$ is the turbulent crossing time on scale $\lambda\sim 1/k$, and the turbulence dispersion obeys $v_t^2(\lambda)\propto \lambda^{p-1}$ thus $\tau_{\lambda} \propto \lambda^{\frac{p-3}{2}}$ (in our simulations we use $p=2$, appropriate for supersonic turbulence, see \citep{Murray_supersonic_Burgers_analytic, Schmidt_supersonic_sim, Federrath_slope}).

\subsubsection{The Equation of State}

It is easy to convince oneself that a purely isothermal or polytropic equation of state (EOS) would be a very poor description of the complex physical processes contributing to the cooling and heating of clouds, however, modeling these processes in detail would require full numerical simulations. Instead we try to find a simple, heuristic EOS that captures the behaviors critical to our calculation. One of the most important effect during collapse is the transitioning from the state where the cooling radiation efficiently escapes from the cloud to the state where the cloud becomes optically thick to it and heats up as it contracts. As the virial parameter is assumed to be constant, this leads to a decrease in turbulence, which effectively arrests fragmentation. This is essential to reproduce the IMF shape as pure isothermal collapse would lead to an infinite fragmentation cascade. We adopt the same effective polytropic EOS model as \CMFIMF where for small time steps (compared to the dynamical time):
\be
\label{eq:T_evol}
T(\xvect,t+\Delta t)=T(\xvect,t)\left(\frac{\rho(\xvect,t+\Delta t)}{\rho(\xvect,t)}\right)^{\gamma\left(t\right)-1},
\ee
where $\gamma\left(t\right)$ is the effective polytropic index of the cloud at time $t$.

One of the main goals and advantages of our framework is that it allows the exploration of different physical EOS models simply and efficiently. For example, let us consider first the volume-density ($n$) dependent EOS model based on works like \citealt{Masunaga_EQS_highgamma_ref, Glover_EQS_lowgamma_ref} that follows the form
\be
\label{eq:gamma_trad}
\gamma(n) =\begin{cases} 0.8 &n < 10^5\, \rm{cm^{-3}} \\
															1.0 &  10^5< \frac{n}{\rm{cm^{-3}}} < 10^{10}   \\
															1.4 &n > 10^{10}\, \rm{cm^{-3}} \\
		  \end{cases}.
\ee
Simulations have shown that this leads to a 'turnover' only at extremely low masses ($\sim 0.001\,\solarmass$, Fig. \ref{fig:IMF_timeavg} later), making the IMF nearly a pure power-law at the observable masses. We will explore model and some of its physical consequences for observables in more detail in a future paper, but explicitly show below that our semi-analytic model also captures this behavior. This is a valuable vindication both of the accuracy of the semi-analytic model (compared to full numerical simulations), and of the need for additional physics to establish the turn-over of the IMF. 

For purposes of this study, let us assume that we do not know the detailed origin of such physics (it may be due to magnetic fields, or radiative heating, for example, both of which we will explore in detail in follow up papers). The simplest approach, and one commonly adopted in numerical simulations, is to parametrize their effects via an 'effective equation of state'. Motivated by the work on radiative feedback from [(\citealt{Bate_2009_rad_importance, Krumholz_stellar_mass_origin}), let us consider a toy model where the effective EOS is not volume-density but surface-density ($\Sigma$) dependent: 
\be
\label{eq:gamma}
\gamma(\Sigma) =\begin{cases} 0.7 &\Sigma < 3\, \solarmass/\pc^2 \\
															0.094 \ln{\left(\frac{\Sigma}{3\,\solarmass/\pc^2}\right)}+0.7 & 3 < \frac{\Sigma}{\solarmass/\pc^2} < 5000   \\
															1.4 &\Sigma > 5000\, \solarmass/\pc^2 \\
		  \end{cases}.
\ee
This is the same EOS as we used in \CMFIMF. Note that the \myquote{turnover} where this becomes \myquote{stiff} is at much lower surface densities than we would obtain if we modeled cooling physics alone (\citealt{Glover_EQS_lowgamma_ref}) which would essentially give the same answer as our $\gamma(n)$ case above (for a comparison of the two types of EOS models, see \cite{guszejnov_feedback_necessity}. Instead, we are assuming some form of physics makes the EOS stiffen at much higher surface densities -- we choose the particular value here {\em empirically}, because it provides a reasonable fit to the observed IMF. We will then explore the consequences of such a parametrization, for the IMF and its time-evolution in different clouds
	
\subsection{Collapse: criterion and evolution}

It has been shown in \CMFIMF and \genturb that the critical density for a (compared to the galactic disk) small, homogeneous, spherical region of radius $R$ to become self gravitating is 
	\be
\frac{\rho_{\rm{crit}}(R)}{\rho_0}=\frac{1}{1+\mach_{\rm{edge}}^2}\lbra\frac{R}{R_0}\rbra^{-2}\left[\left(\frac{T(R)}{T_0}\right)+\mach_{\rm{edge}}^2\lbra\frac{R}{R_0}\rbra^{p-1}\right],
\label{eq:collapse_threshold_T}
\ee
where the two terms represent thermal and turbulent energy respectively. $T(\lambda)$ is the temperature averaged over the scale $\lambda$, while $T_0$ is the mean temperature of the whole collapsing cloud and we used the following scaling of the turbulent velocity dispersion and Mach number $\mach$
	\be
	\label{eq:mach_scaling}
	\mach^2(R)\equiv\frac{v_t^2(R)}{\lang c_s^2\left(\rho_0\right)\rang}=\mach^2_{\rm edge}\left(\frac{R}{R_0}\right)^{p-1},
	\ee
	 where $R_0$ is the size of the self gravitating parent cloud and $p$ is the turbulent spectra index, so the turbulent kinetic energy scales as $E(R)\propto R^{p}$; generally $p\in [5/3;2]$, but in this paper, just like in \CMFIMF we assume $p=2$ as is appropriate for supersonic turbulence.
	
	It should be noted that the fragmentation process is complex even in the idealized case of homologous collapse (see \citealt{Hanawa_bar_perturbation, Ntormousi_core_shell_instability}). This means that our method of finding self gravitating subregions using Eq. \ref{eq:collapse_threshold_T} is a strong approximation, however, a proper treatment would require drastically more computation power which would go against one of the primary goal of the framework: the rapid exploration of parameter space and testing of physical models. 

Our goal is to create a model that resolves clouds from GMC to protostellar scales, so the initial structures of the model are the GMCs which themselves are self gravitating (first crossing scale in the excursion set formalism). This means they must satisfy Eq. \ref{eq:collapse_threshold_T}, which for spherical clouds ($M(R)=(4\pi/3)\, R^3\,\rho(R)$) in isothermal parents yields the mass-size relation:
\be
M=\frac{M_{\rm sonic}}{2}\frac{R}{R_{\rm sonic}}\left(1+\frac{R}{R_{\rm sonic}}\right).
\label{eq:mass_size}
\ee
Note that for very high mass clouds a correction containing the angular frequency of the galactic disk would appear, however this term is small (see \genturb for details). Eq. \ref{eq:mass_size} introduces $R_{\rm sonic}$ which is the sonic length, the scale on which the turbulent velocity dispersion is equal to the sound speed, so in an isothermal cloud using the scaling of Eq. \ref{eq:mach_scaling}, we expect
\be
\label{eq:sonic_length}
R_{\rm sonic}=R_0\mach_{\rm{edge}}^{-2/(p-1)}.
\ee
Meanwhile $M_{\rm sonic}$ is defined as the minimum mass required for a sphere with $R_{\rm sonic}$ radius to start collapsing so
\be
\label{eq:sonic_mass}
M_{\rm sonic}=\frac{2}{Q_{\rm coll}}\frac{c_s^2 R_{\rm sonic}}{G},
\ee
where $G$ is the gravitational constant and $Q_{\rm coll}$ is the virial parameter for a sphere of the critical mass for collapse (see Eq. \ref{eq:virial} later). For reasonable galactic parameters and temperatures $R_{\rm sonic}\approx 0.1\,\pc$ and $M_{\rm sonic}\approx 6.5\,\solarmass$ (assuming we use the value for $Q_{\rm coll}$ we specify in Sec. \ref{sec:collapse_evol}).

Since the GMC in question has just started collapsing, the turbulent velocity at its edge must (initially) obey the turbulent power spectrum. Thus $v_t^2(R)\propto R$ for the supersonic and $v_t^2(R)\propto R^{2/3}$ (the Kolmogorov scaling) for the subsonic case. Using the mass-size relation of Eq. \ref{eq:mass_size} leads to the following fitting function
\be
\frac{\left(1+\mach^2_{\rm edge}\right)\mach^2_{\rm edge}}{1+\mach^{-1}_{\rm edge}}=\frac{M}{M_{\rm sonic}},
\ee
which exhibits scalings of $M\propto\mach^3$ for the subsonic and $M\propto\mach^4$ for the supersonic case respectively, and (coupled to the size-mass relation above) very closely reproduces the observed linewidth-size relations (\citealt{Larson_law, Bolatto_2008, clustering_Lada}). Note that dense regions will deviate from this scaling, as observed (see references above), because collapse ``pumps'' energy into turbulence (\citealt{Brant_turb_pumping, Murray_star_formation, Murray_2015_turb_sim}).

\subsubsection{Evolution of Collapsing Clouds}\label{sec:collapse_evol}

One of the key assumptions of the previous models in \CMFIMF and \genturb is that the kinetic energy of collapse pumps turbulence (\citealt{Brant_turb_pumping, Murray_star_formation, Murray_2015_turb_sim}) whose energy is dissipated on a crossing time. As turbulent motion provides support against collapse, the collapse can only continue after this extra energy has been dissipated by turbulence (see Sec. 9.2 in \genturb for details). This leads to the following equation for the contraction of the cloud:
\be
\frac{\dderiv\tilde{r}}{\dderiv\tilde{\tau}}=-\tilde{r}^{-1/2}\left(1-\frac{1}{1+\mach_{\rm{edge}}^2(\tilde{\tau})}\right)^{3/2},
\label{eq:scale_evol}
\ee
where $\tilde{r}(t)=R(t)/R_0$ is the relative size of the cloud at time $t$ while $\tilde{\tau}\equiv t/t_0$ is time, normalized to the initial cloud dynamical time $t_0\sim 2 Q_{\rm coll}^{-3/2}\left(G M_0/R_0^3\right)^{-1/2}$ (see \genturb for derivation). In this case the initial dynamical time ($t_{0}$) and the crossing time only differ by a freely-defined order unity constant, so in our simulations we consider them to be equal without loss of generality.

The other key assumption of the model is that collapse happens at constant virial parameter. We define $Q_{\rm coll}$ as
\be
Q_{\rm coll} \frac{G M}{R}=c_s^2+v_t^2=c_s^2\left(1+\mach_{\rm{edge}}^2\right).
\label{eq:virial}
\ee
Note that $Q_{\rm coll}$ is not the Toomre Q parameter, merely the ratio of kinetic energy to potential energy needed to destabilize the cloud, thus the higher $Q_{\rm coll}$ the more unstable clouds are to fragmentation. One can find $Q_{\rm coll}$ using the Jeans criterion:
\be
0\geq\omega^2=\left(c_s^2+v_t^2\right)k^2-4\pi G\rho,
\ee
which for the critical case ($\omega=0$) leads to
\be
Q_{\rm coll}=\frac{3}{k^2 R^2}.
\ee
One would be tempted to substitute in $k=2\pi/R$, but that would be incorrect, as we have a spherical overdensity with R radius to which the corresponding sinusoidal wavelength is not R. We therefore chose $k=\frac{\pi}{2 R}$ which yields $Q_{\rm coll}=12/\pi^2\approx 1.2$. Note that all formulas contain $c_s^2/Q_{\rm coll}\propto T/Q_{\rm coll}$ so an uncertainty in the virial parameter is degenerate with an uncertainty in the initial temperature.

Combined, the above equations completely describe the collapse of a spherical cloud, as the EOS (Eq. \ref{eq:T_evol}-\ref{eq:gamma}) sets the temperature and thus the sound speed. Using that, Eq. \ref{eq:virial} provides the edge Mach number, which allows us using Eq. \ref{eq:scale_evol} to calculate the contraction speed.

\subsection{Differences from previous models}\label{sec:model_diff}

So far we are following the same assumptions as \CMFIMF and \genturb, however, instead of simulating a stochastic density field averaged on different scales around a random Lagrangian point (the basis of analytic excursion set models) we use a grid in space and time. This means that we directly evolve the $\delta\lbra k\rbra$ modes to simulate the density field. This allows us to preserve spatial information as we now have information about the relative positions and velocities of substructures.

Having a proper density field not only allows us to take basic geometrical effects into account (as substructures are still assumed to be spherical) but it allows a proper application of the self gravitation condition of Eq. \ref{eq:collapse_threshold_T}. The excursion set formalism finds the smallest self gravitating structure a point is embedded in. The problem is that this ``last crossing'' structure may have further self gravitating fragments which do not contain the aforementioned point. These substructures will form protostars of their own (see Fig. \ref{fig:fragment_fig}) leaving their parent cloud with less mass which in turn might not be self gravitating anymore. This is not addressed in excursion set models which instead simply assume 100\% of the mass ending up in protostars of different sizes (which of course is not realistic), while the proposed grid model predicts only about 5\% (see Sec.\ref{sec:IMF}), which in fact depends on the physical assumptions of the model (i.e. how to deal with unbound material).

It should be noted that like the model of \CMFIMF, in this first study we include no explicit feedback mechanism. Instead the model utilizes a few crude approximations to account for the qualitative effects of feedback. First, it is assumed that the clouds that becomes unbound by fragmentation stop collapsing and ``linger'' for a few dynamical times (during which they may form new self gravitating fragments) before being heated up/blown up/disrupted by feedback from the newly created protostars in such a fashion that they can no longer participate in star formation\footnote{For example photoionization can destroy the molecular cloud (\citealt{Dale_ion_feedback, Walch_ion_feedback, Geen_massive_feedback}), while both supernovae (\citealt{Iffrig_SN_feedback}) and outflows (\citealt{Arce_outflow_review}) can provide momentum for turbulence or eject material.}. Note that this assumption is made for convenience, it is not inherent in the code as it is possible to implement direct feedback prescriptions. Similarly magnetic fields are neglected in this base model, but can be easily implemented into the framework. Like in \CMFIMF we neglected the effects of accretion and protostellar fragmentation when comparing to the IMF as the protostellar system mass function (from now on \textit{PSMF}) is already a good enough qualitative fit so their effects are assumed to be modest (except for the very high and low mass ends where fragmentation could provide a high mass cut off while accretion could affect the turnover point, see \citealt{Offner_protostar_MF} for details on the protostellar mass function). We would also like to note that it is possible to apply a crude implementation of supernova feedback by simply stopping the evolution after a few Myrs (when enough supernovae have exploded to unbind the GMC). Since the simulation provides a time dependent output, it can be done during post-processing. Of course, the point of our framework is that one could easily add models for feedback, and/or accretion if desired.

We would like to note that using hydrodynamical simulations would allow a much more realistic treatment of certain details of the problem, however the large dynamic range ($10^{-5}-100\,\pc$) and the long range gravitational interactions make such attempts extremely computationally intensive, preventing one from getting substantial statistics. A further issue with direct hydrodynamical simulations is that they involve the full, detailed form of all physical interactions, making it harder to pinpoint the primary driving mechanisms behind certain phenomena.

In summary we propose a semi-analytical model which has negligible computational cost but still captures phenomena (e.g. spatial correlation, motion of objects, complicated time dependence) which are beyond the capabilities of the analytical excursion set formalism. Our intention in this paper is not to present a \myquote{complete} model of star formation, but rather to illustrate the power of this approach with a first study involving only turbulence and self-gravity.

\section{Evolution of the IMF and CMF in GMCs}\label{sec:imf_cmf_evol}

\begin{figure}
\begin {center}
\includegraphics[width=\linewidth]{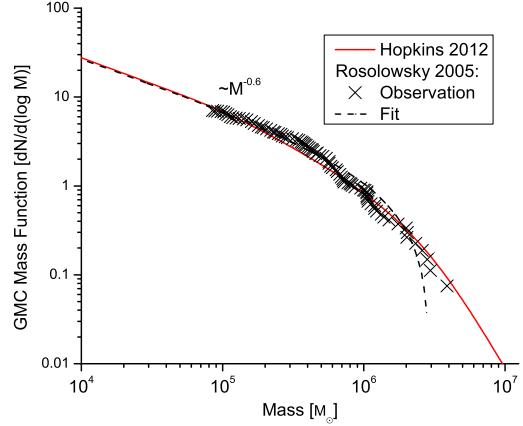}
\caption{Initial mass function of GMCs according to the excursion set model of \protect\cite{core_IMF} compared to the observations (X symbols) and empirical fitting function (dashed black line) of \protect\cite{Rosolowsky_2005_GMC}. The normalization of the plot is arbitrary.}\label{fig:GMC_MF}
\end {center}
\end{figure}

In this section we present an application of the model for simulating the collapse of an ensemble of GMCs (distributed following the first crossing mass function obtained by \citealt{core_IMF}, see Fig. \ref{fig:GMC_MF}). This includes simulating a number of GMCs of different masses where the initial conditions are set by Eq. \ref{eq:mach_scaling} and Eq. \ref{eq:mass_size}. The clouds are assumed to start with fully formed turbulence (as GMCs form out of an already turbulent medium) which means that before simulating the collapse the density field is initialized to have the appropriate lognormal distribution. The output of the code contains the formation time and properties (e.g. mass, position, velocity) of individual protostars along with snapshots of the hierarchical structure of bound objects at different times. In Sec. \ref{sec:CMF} we investigate the latter and compare the distribution of nonfragmented structures with the observed CMF. Later, in Sec. \ref{sec:IMF} we discuss the time evolution of PSMF and how it relates to the IMF and whether it can be universal without invoking feedback physics.

\subsection{Fragmentation and self-gravitating substructures: the observed CMF}\label{sec:CMF}

\begin{figure}
\begin {center}
\includegraphics[width=\linewidth]{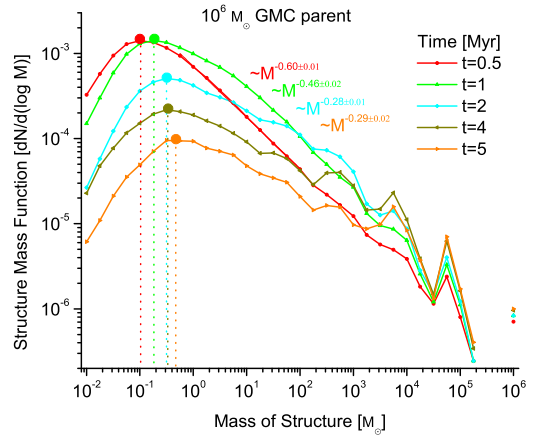}
\caption{Time evolution of number of bound structures of different masses in a parent GMC of $10^6\,\solarmass$. Here we count \emph{all} self-gravitating structures, including clouds embedded in other clouds, cores etc. The plot is normalized so that integrated mass ($\int{M \frac{\dderiv N}{\dderiv \log{M}} \dderiv \log M}$) corresponds to the mass of gas bound in self gravitating clouds relative to the total mass of the parent GMC, which explains the decreasing trend with time as more and more gas ends up in either protostars or becomes unbound. The upper end cuts off close to the parent GMC mass. The high mass power law fitting is done according to Footnote \ref{foot:fitting}.}
\label{fig:bound_struct_evol}
\end {center}
\end{figure}

It is well known that during their collapse clouds fragment into smaller self gravitating structures (see Fig. \ref{fig:fragment_fig}). It is instructive to see how much mass is bound in structures of different sizes. Fig. \ref{fig:bound_struct_evol} shows the time evolution of the number of structures of different sizes counting all ``clouds-in-clouds'', which follows a distribution similar to the observed IMF and CMF (for quick overview see \citealt{IMF_universality}), however it has a significantly shallower slope\footnote{\label{foot:fitting}In this paper the approximate high mass end behavior is estimated by fitting a power law between $0.5\,\solarmass$-$100\,\solarmass$. The error presented in the figures only account for the uncertainty in the fitting.} of roughly $M^{-0.3}$. The distribution is established fairly quickly and is maintained until the collapse of the parent cloud ends. This mass function of bound structures is consistent with the cloud in cloud picture shown in Fig. \ref{fig:fragment_fig} in that there is a vast hierarchy of bound structures embedded in each other.

\begin{figure}
\begin {center}
\includegraphics[width=\linewidth]{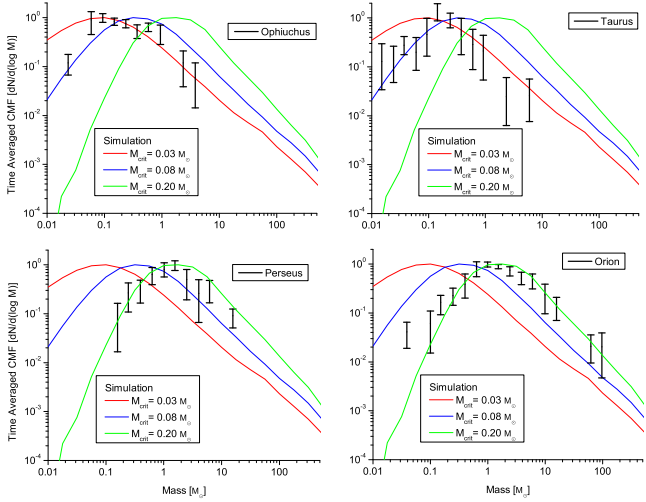}
\caption{Comparison of the average simulated CMF with the observed CMF by \protect\cite{Sadavoy_observed_CMF} in different clouds in the Milky Way (the plot is normalized so that the peak of the CMF is set to unity). Note that observations which are below the completeness limit are also included (see the original paper for details). The simulated CMFs are averaged both over time (assuming the age of GMCs is uniformly distributed in the [0,5] Myr range) and the GMC mass function (following Fig. \ref{fig:GMC_MF}). The different initial critical masses in this case reflect having different $T/Q_{\rm coll}$ values, for definition see Eq. \ref{eq:Mcrit_opacity_simple}.}
\label{fig:nonfragmenting_CMF}
\end {center}
\end{figure}

Observationally finding the substructure of a GMC is very challenging (although see \citealt{Rosolowsky2008_dendogram}), most observers instead concentrate on the so called \textit{cores} which are collapsing clouds that have no self gravitating fragments. Figure \ref{fig:nonfragmenting_CMF} shows the total CMF (time and mass averaged over an ensemble of GMCs following the distribution shown in Fig. \ref{fig:GMC_MF}) for different inital parameters. The simulated CMF reproduces the shape of observed results, having both a turnover point and a slightly shallower high mass slope ($\sim M^{-1.15}$) than the canonical Salpeter result of $\sim M^{-1.35}$ for the IMF (see \citealt{IMF_universality}).

\begin{figure}
\begin {center}
\includegraphics[width=\linewidth]{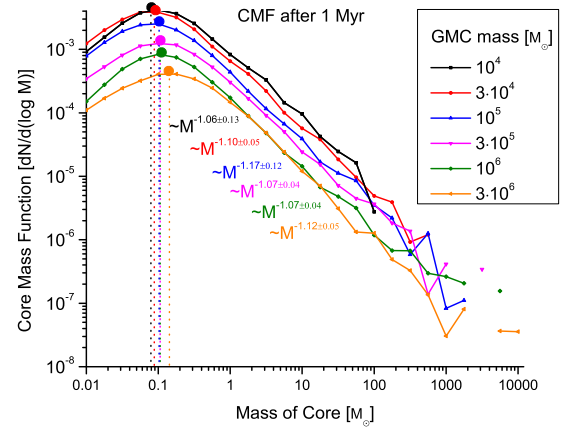}
\caption{The CMF in GMCs of different masses 1 Myr after collapse starts for each cloud (using EOS of Eq. \ref{eq:gamma}). The plot is normalized so that integrated mass corresponds to the relative mass of gas bound in cores, the peaks are denoted with solid circles. The high mass power law fitting is done according to Footnote \ref{foot:fitting}. Both the turnover mass and the high mass slope exhibit very little sensitivity to the mass of the parent GMC similar to what was found by \protect \cite{HC08, Hennebelle2012}.}\label{fig:nonfragment_after1Myr}
\end {center}
\end{figure}

\begin{figure*}
\includegraphics[width=0.45\linewidth]{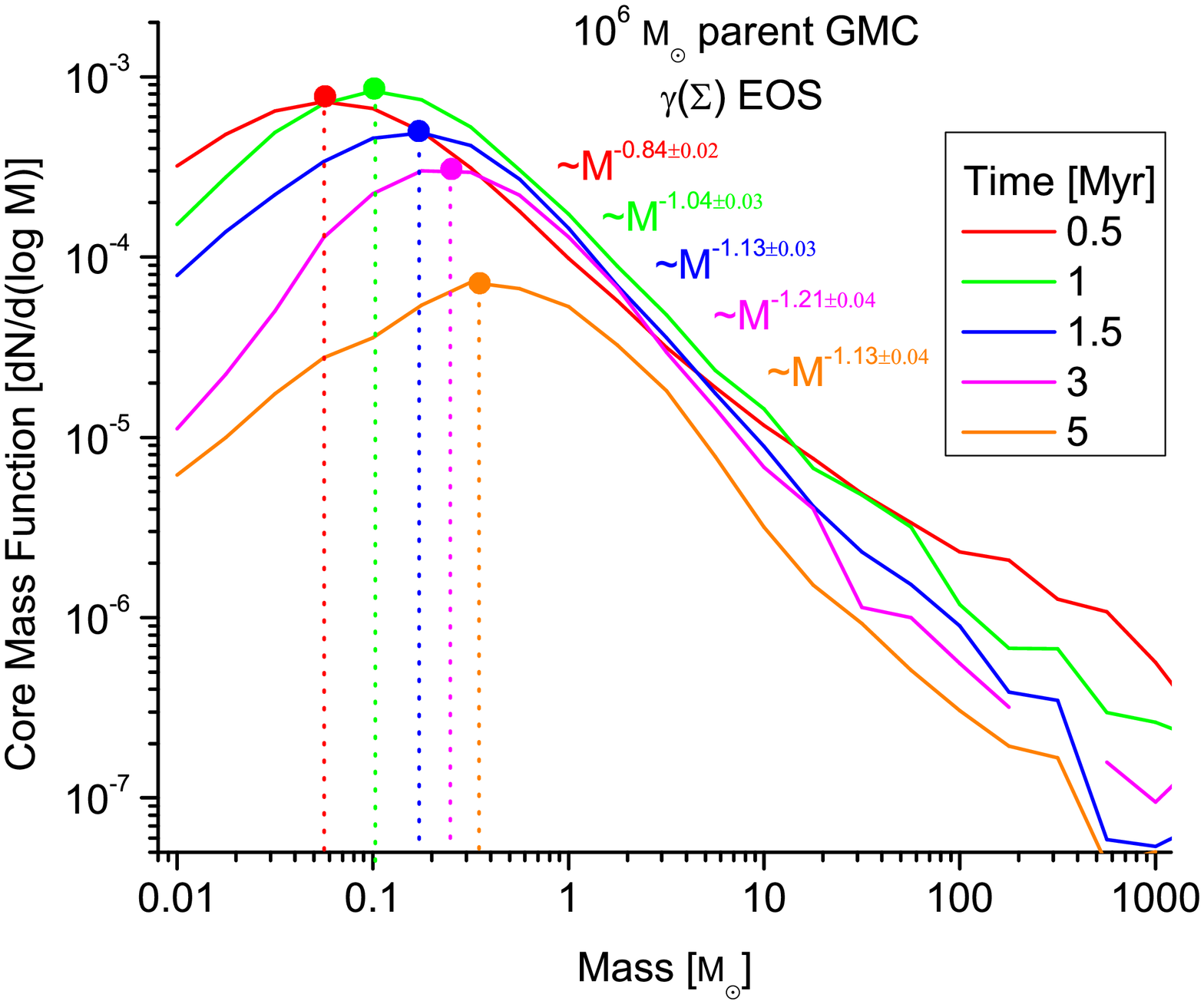}
\includegraphics[width=0.45\linewidth]{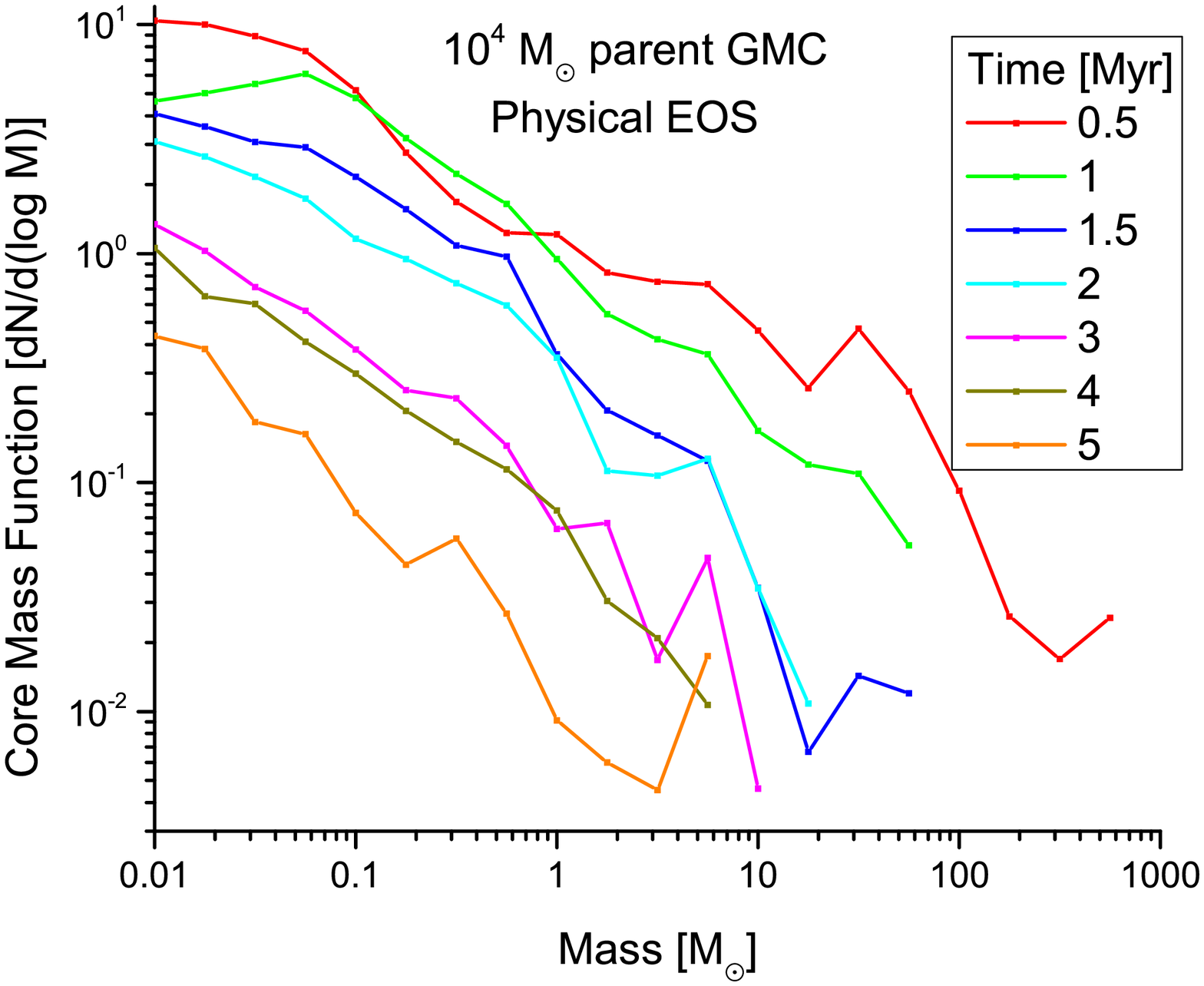}
\caption{\textbf{Left}: Time evolution of the CMF in a $10^6\,\solarmass$ parent GMC using the $\gamma(\Sigma)$ EOS of Eq. \ref{eq:gamma}. The plot is normalized so that integrated mass corresponds to the mass of gas bound in self gravitating clouds relative to the total mass of the parent GMC, which explains the downwards trend since less and less gas is bound in cores as more protostars are produced and the cloud gets heated by contraction. The high mass power law fitting is done according to Footnote \ref{foot:fitting}. There is a clear trend in the turnover mass (the peaks are denoted with solid circles) which increases significantly while preserving the overall shape of the function (e.g. high mass slope). \textbf{Right}: Time evolution of the CMF in a $10^4\,\solarmass$ parent GMC using the physically motivated EOS of Eq. \ref{eq:gamma_trad} (a density dependent EOS where the transition point to the $\gamma>1$ regime is calculated from cooling physics). As expected the CMF has a peak around the sonic mass at early times, however, that feature gets \myquote{washed out} by the fragmentation cascade which is not arrested by this EOS until very small scales.}\label{fig:nonfragment_in_parent_evolve}
\end{figure*}

Fig. \ref{fig:nonfragment_after1Myr} clearly shows that there is very small difference between the CMF turnover masses and high mass slopes between GMCs of different sizes after 1 Myr. This is because early collapse is roughly isothermal so these clouds all have the same characteristic fragment mass ($M_{\rm crit}$, see Eq. \ref{eq:Mcrit_opacity_simple} for details). Systems which are on the same linewidth-size relation (i.e. they form out of the same turbulent cascade) will always have the same $M_{\rm sonic}$, $M_{\rm crit}$ (see \citealt{HC08, core_IMF}). During later evolution the GMCs heat up at a different pace as the dynamical times are different. Meanwhile Fig. \ref{fig:nonfragment_in_parent_evolve} shows that  there is a clear trend of increasing turnover mass with time in each cloud. This phenomenon and its possible cause is further investigated in Sec. \ref{sec:IMF}. This trend is not visible in case of the physical EOS of Eq. \ref{eq:gamma_trad} as the peak is well below the stellar mass scales (see Fig. \ref{fig:IMF_timeavg}). Nevertheless, this scenario shows that in the absence of a dominant $M_{\rm crit}$ the initial CMF turns over around the sonic mass scale (as shown by previous analytical works e.g. \citealt{HC08, core_IMF}), but this mass scale gets \myquote{forgotten} during the fragmentation cascade. 

\subsection{Evolution of the PSMF}\label{sec:IMF}

We now examine the mass function of the final collapsed objects, the protostellar system mass function (PSMF).

\begin{figure}
\begin {center}
\includegraphics[width=\linewidth]{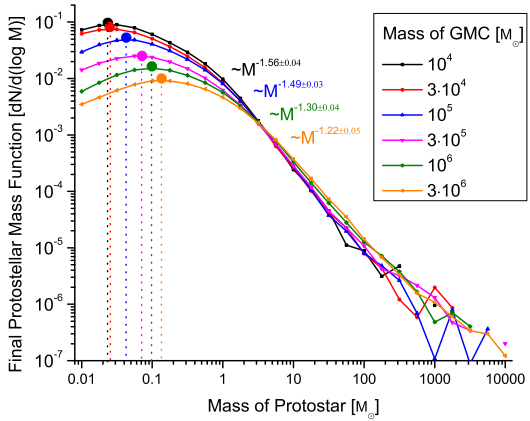}
\caption{Protostellar system mass function (PSMF) after collapse ends in parents of different masses assuming our simple equation of state. The Salpeter slope is always present (the high mass power law fitting is done according to Footnote \ref{foot:fitting}). For these assumptions there appears to be ``too many'' brown dwarfs, and too much dependence on the parent GMC mass. These are the direct consequences of the EOS of the gas.}\label{fig:GMC_starnumber_dist}
\end {center}
\end{figure}

In Fig. \ref{fig:GMC_starnumber_dist} we show that parent clouds of all masses produce similar to Salpeter scalings the high mass end with lower mass clouds producing slightly steeper slopes. Also, there is a clear trend of increasing turnover mass with increasing parent mass, unlike the case of the CMF (See Fig. \ref{fig:nonfragment_after1Myr}). It is worth noting that the GMC mass function is top heavy, which means that the high mass clouds dominate the integrated mass function. If we accept this result then it suggests a possible observational bias of the IMF as most observations focus on smaller clouds in the Milky Way. Also, turbulent fragmentation does not produce a cloud mass dependent \myquote{maximum stellar mass}.

The increasing turnover mass for both PSMF and CMF is related to the equation of state. In a turbulent cloud, self gravitating fragments of different sizes form, which (according to the EOS of Eq. \ref{eq:gamma}) have different effective polytropic indices. According to the EOS there exists a threshold in the surface density ($\Sigma_{\rm crit}$) above which $\gamma>4/3$, stabilizing the cloud against further fragmentation. Thus it is instructive to find the critical mass ($M_{\rm crit}$) corresponding to $\Sigma_{\rm crit}$. Using the collapse condition of Eq. \ref{eq:collapse_threshold_T} and expanding up to linear order in $\gamma$ around 1 (this is a good approximation during most of the cloud's lifetime as the collapse starts at close to isothermal conditions) yields that $\Sigma>\Sigma_{\rm crit}$ requires that
\be
R<R_{\rm crit}=R_0\frac{\gamma \lbra\frac{\Sigma_{\rm crit}}{\Sigma_0}\rbra^{\gamma-1}}{\frac{\Sigma_{\rm crit}}{\Sigma_0}\lbra 1+\mach_{\rm edge}^2\rbra-\mach_{\rm edge}^2+\gamma-1},
\label{eq:Rcrit_opacity}
\ee
where $R$ is the fragment radius and $R_0$, $\Sigma_0$, $\gamma=\gamma\lbra\Sigma_0\rbra$ are the radius, surface density and the effective polytropic index of the parent cloud. From Eq. \ref{eq:Rcrit_opacity} we can find the critical mass $M_{\rm crit}=4\pi R^2 \Sigma_{\rm crit}$ below which fragments are unlikely to collapse (note: according to the EOS of Eq. \ref{eq:gamma} the critical surface density $\Sigma_{\rm crit}\approx 2400\,\solarmass/\pc^2$). These formulas can be simplified by assuming isothermal collapse ($\gamma\simeq 1$) and that the parent GMC is highly supersonic ($\mach_{\rm edge}^2\gg 1$), Eq. \ref{eq:sonic_length} yields then:
\be
R_{\rm crit}\approx\frac{R_0 \Sigma_0}{\mach_{\rm edge}^2\Sigma_{\rm crit}}=R_{\rm sonic} \frac{\Sigma_0}{\Sigma_{\rm crit}}.
\ee
Using the mass-size relation of Eq. \ref{eq:mass_size} and that $R_0\gg R_{\rm sonic}$ we obtain
\be
M_{\rm crit}\approx \frac{4\pi R_{\rm sonic}^2 \Sigma_0^2}{\Sigma_{\rm crit}}=\frac{M_{\rm sonic}^2}{16\pi R_{\rm sonic}^2\Sigma_{\rm crit}}=\frac{c_s^4}{4\pi G^2 Q_{\rm coll}^2\Sigma_{\rm crit}}\propto \frac{T^2}{\Sigma_{\rm crit}}.
\label{eq:Mcrit_opacity_simple}
\ee
The critical mass only depends on the cloud temperature and the equation of state. A similar sensitivity to the initial temperature has been found by \cite{Bate_2009_rad_importance} using a Jeans mass argument. Assuming that there exists a critical density $\rho_{\rm crit}$ where some physics terminates the fragmentation cascade the corresponding Jeans mass will simply be $\propto T^{3/2}$. It is easy to see that this is the same result one would get when trying to find the critical mass using a $\gamma(n)$ EOS.

\begin{figure}
\begin {center}
\includegraphics[width=\linewidth]{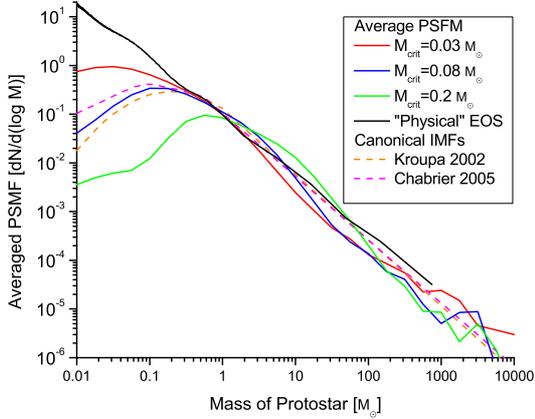}
\caption{Evolution of the averaged PSMF (normalized to integrated mass) for different initial critical masses (set by having different $T/Q_{\rm coll}\Sigma_{\rm crit}$ values, for definition see Eq. \ref{eq:Mcrit_opacity_simple}) compared to results using the \myquote{traditional} EOS of Eq. \ref{eq:gamma_trad} and the canonical IMF of \protect\cite{Kroupa_IMF} and \protect\cite{Chabrier_IMF}. The PSMF is averaged both over time (assuming the age of GMCs is uniformly distributed in the [0,5] Myr range) and the GMC mass function (following Fig. \ref{fig:GMC_MF}). We included the standard $M_{\rm crit}=0.03\,\solarmass$ (solid red), an $M_{\rm crit}=0.08\,\solarmass$ (solid blue) and an $M_{\rm crit}=0.2\,\solarmass$ (solid black) scenarios with the $\gamma(\Sigma)$ EOS along with a run which had the physically motivated $\gamma(n)$ EOS of Eq. \ref{eq:gamma_trad}. For realistic temperatures ($10-30\,\rm{K}$) the critical mass of the latter is well below the stellar mass range so the PSMF becomes a pure power law. Meanwhile, for the $\gamma(\Sigma)$ EOS case the PSMF shape is similar for different critical masses, and there is a clear shift of the peak to higher masses with increasing $M_{\rm crit}$. In all cases the high mass end is close to the Salpeter result.}
\label{fig:IMF_timeavg}
\end {center}
\end{figure}

Fig.\ref{fig:IMF_timeavg} shows the time evolution of the time and ensemble averaged PSMF for different initial $M_{\rm crit}$ values (the different critical masses in these cases arise from having different $\sigma/Q_{\rm coll}\Sigma_{\rm crit}$; where we fix $Q_{\rm coll}$ and $\Sigma_{\rm crit}$ and vary $T_{\rm init}$, for definition see Eq. \ref{eq:Mcrit_opacity_simple}) which all produce a shape similar to the IMF but with different peak masses. If we compare the results to the canonical IMF fitting functions of \cite{Kroupa_IMF} and \cite{Chabrier_IMF}, then it is clear that the average PSMF always reproduces the Salpeter scalings however the turnover point is heavily influenced by $T/Q_{\rm coll}\Sigma_{\rm crit}$ . Since $Q_{\rm coll}$ is a constant this implies that the average temperature of the cloud could have a significant effect on the turnover point if $\Sigma_{\rm crit}$ is constant. Meanwhile, Fig.\ref{fig:IMF_timeavg} also shows that the physical EOS of Eq. \ref{eq:gamma_trad} has such a low characteristic mass that the resulting PSMF in the stellar mass range is just a power law. Nevertheless, the position of the peak is still sensitive to the initial conditions ($\propto T^{3/2}$), if one extends the plot to substellar mass scales.

\begin{figure}
\begin {center}
\includegraphics[width=\linewidth]{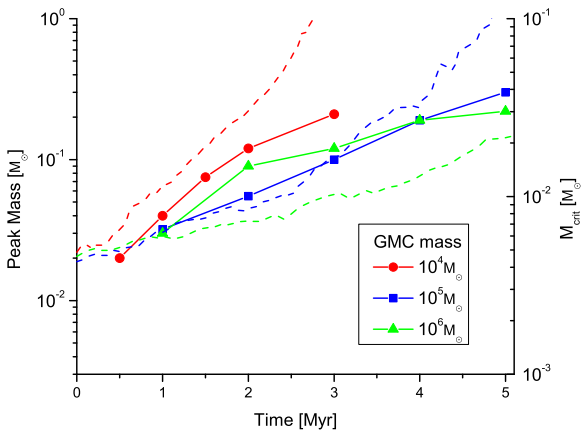}
\caption{The peak masses of the PSMF of different time intervals (solid line with symbols) and the critical mass (dashed lines) for different parent GMC masses according to Eq. \ref{eq:Rcrit_opacity}. The critical mass correctly predicts the qualitative evolution of the peak mass.}\label{fig:OpacityMassVSPeakMass}
\end {center}
\end{figure}

Fig. \ref{fig:OpacityMassVSPeakMass} shows how this critical mass evolves in time for our default model assumptions ($\Sigma_{\rm crit}=\rm{const.}$). It is clear that $M_{\rm crit}$ correlates well with the peaks of the PSMF of the corresponding time interval.

\begin{figure}
\begin {center}
\includegraphics[width=\linewidth]{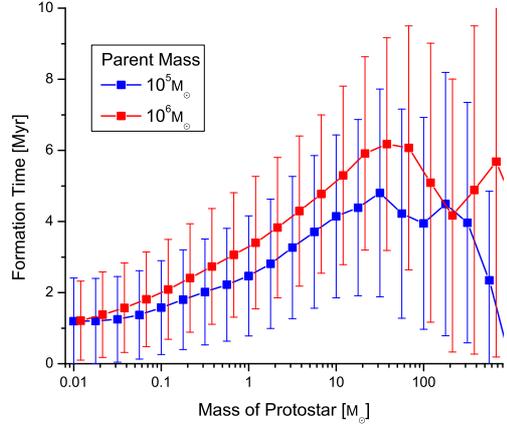}
\caption{Average time of formation for protostars of different masses (the error bars represent the standard deviation) in a model with an \emph{invariant} EOS. There is a clear trend of more massive protostars forming at later times (which is consistent with the shifting of the turnover mass in Fig. \ref{fig:OpacityMassVSPeakMass}), however the scatter is comparable to this difference. Nevertheless it is clear that most massive stars only start forming after roughly a Myr after the cloud starts collapsing. Changing this requires additional physics beyond turbulence, gravity and cooling.}\label{fig:formation_time}
\end {center}
\end{figure}

This increase of the critical mass with time has an interesting consequence. Fig. \ref{fig:formation_time} shows that the average time of formation monotonically increases with the protostellar system mass.

So, if the equation of state does not depend on temperature (e.g. our $\gamma(\Sigma)$ is invariant) then the turnover mass shows a strong ($\propto T^2$) dependence on the initial conditions which would likely lead to a non-universal IMF ($\propto T^{3/2}$ in the $\gamma(n)$ case). A possible solution to this issue is if $\Sigma_{\rm crit}$ from Eq. \ref{eq:Mcrit_opacity_simple} has a temperature dependence. This perfectly plausible, just recall that the effective EOS is just a crude approximation of complex cooling physics. \cite{Bate_2009_rad_importance} argues that radiative feedback effectively weakens the dependence of the Jeans mass on density, making the turnover mass less sensitive to initial conditions. A similar example is provided by \cite{Krumholz_stellar_mass_origin}, where the initially formed protostar 'seed' heats up its environment, preventing it from collapsing. This dense cloud is heated up to $T_{\rm heating}\propto M^{3/8}R^{-7/8}\approx \Sigma^{3/8}$ by the accretion luminosity from the protostar\footnote{One can derive this temperature by assuming an optically thick cloud in equilibrium that is heated by accretion luminosity $L_{\rm acc}\sim \dot{M} \Psi\sim M/t_{\rm ff} \Psi \propto M^{3/2}R^{-3/2}$ and cooled by thermal radiation $L_{\rm cool}\sim 4\pi R^2\sigma_{\rm SB} T_{\rm heat}^4$.}, which, using our EOS language, roughly translates to $\Sigma_{\rm crit}\propto T^2$ which would produce a constant $M_{\rm crit}$, and thus a universal IMF.

In a paper in preparation we will explore this feedback model in a fully spatially-dependent framework. For now, let us consider a simple experiment where $\Sigma_{\rm crit}\propto T^2$.

\begin{figure*}
\begin {center}
\includegraphics[width=0.45\linewidth]{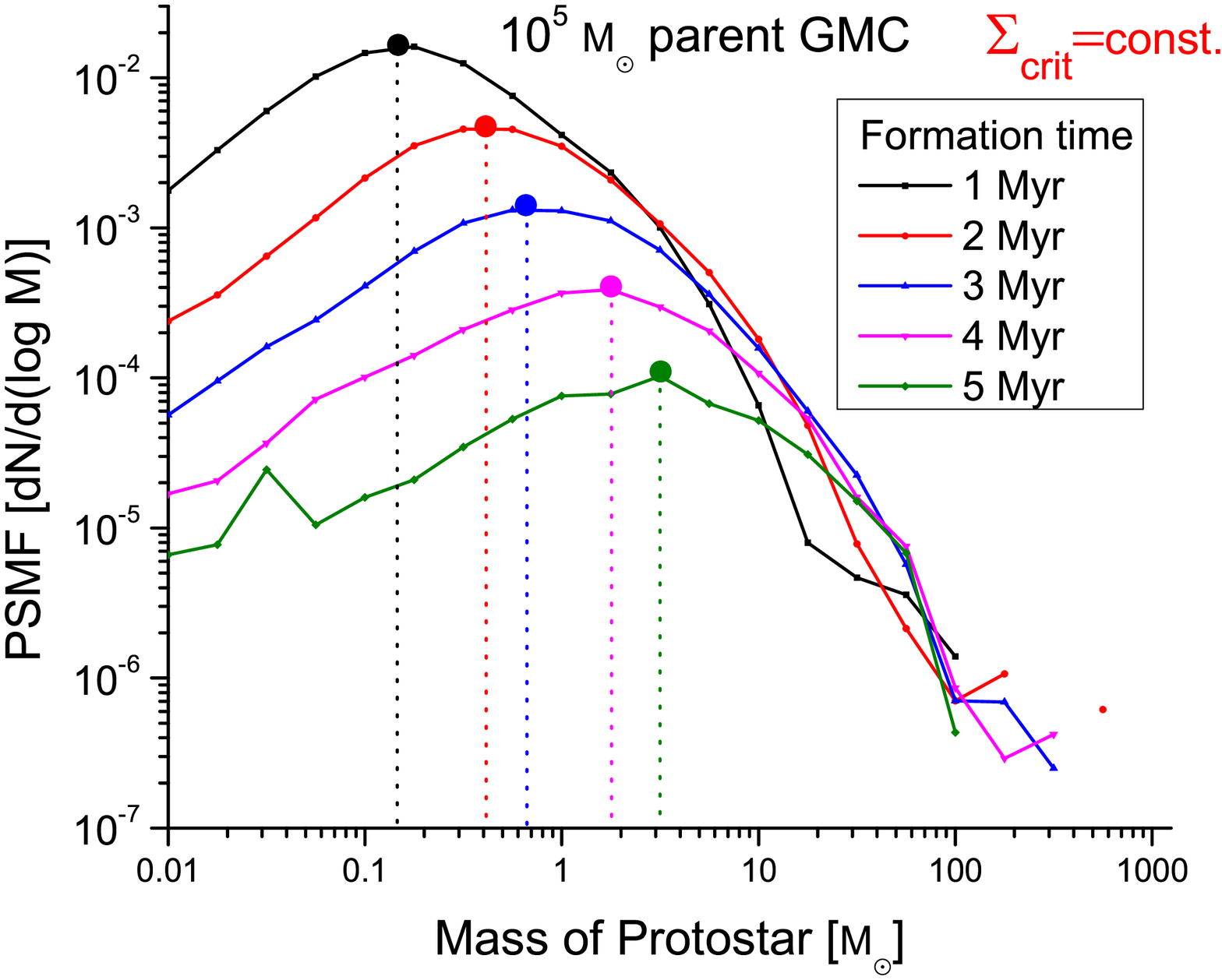}
\includegraphics[width=0.45\linewidth]{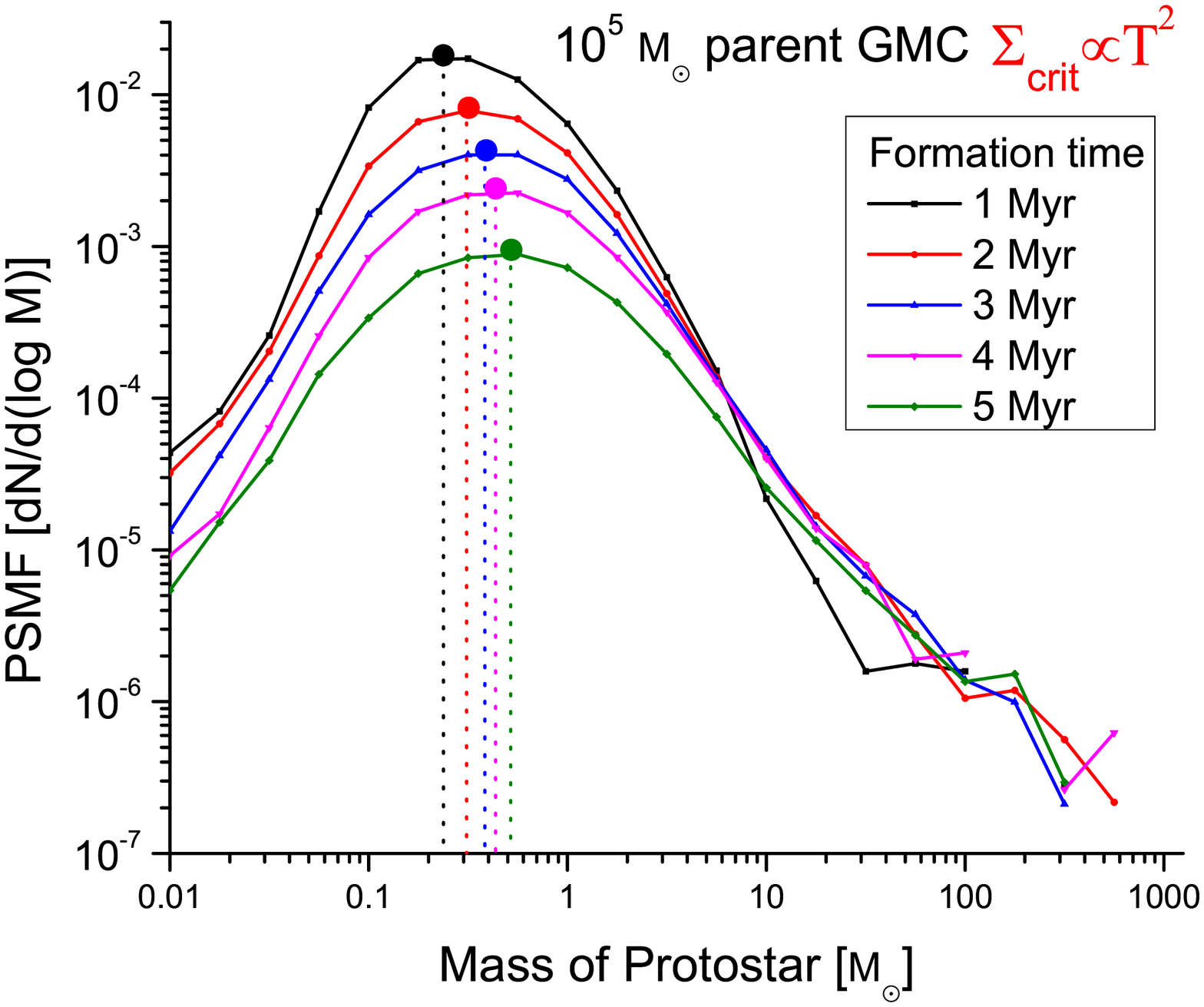}
\caption{PSMF for protostars in a parent GMC of $10^5\,\solarmass$ for an EOS with $\Sigma_{\rm crit}=\rm{const.}$ (left) and for an EOS with $\Sigma_{\rm crit}\propto T^2$ (right). The solid circles show the peaks, which move considerably less for the $\Sigma_{\rm crit}\propto T^2$ case. As implied by Eq. \ref{eq:Mcrit_opacity_simple}, if $\Sigma_{\rm crit}\propto T^2$ then $M_{\rm crit}\sim \rm{const}$, and the IMF becomes invariant.}
\label{fig:compare_PSMF_sigmacrit}
\end {center}
\end{figure*}

Fig. \ref{fig:compare_PSMF_sigmacrit} compares the results of two simulations, one with $\Sigma_{\rm crit}=\rm{const.}$ and one with $\Sigma_{\rm crit}\propto T^2$. Although the latter still shows some time dependence, the shifting of the peak is greatly reduced, making it more consistent with observations, even though the only assumption about feedback was that it prevents collapsed cores from accreting from their surroundings. Note that our aim with this experiment was only to demonstrate what would be required from a purely EOS based model to produce an invariant IMF, any other physics that sets the critical mass of the EOS constant would achieve similar results.

An important question of star formation is what fraction of the gas ends up in stars. The analytical excursion set models like in \CMFIMF could not answer that question as they assume by default that 100\% of the mass ends up in bound structures similar to the Press-Schecter model (\citealt{PressSchechter}) of dark matter halos which they are based on). However, our semi-analytic framework here allows us to explore different assumptions for the time-dependent behavior of both bound and unbound gas, and thus (in principle) to make predictions for this quantity. 

In the \myquote{basic} models presented in this paper, we assume that whenever a core collapses and forms a star, any remaining mass in its parent cloud which is no longer self-gravitating (once the core is fully collapsed) is simply thrown out of the system. This is meant to represent a very crude toy model for the effects of feedback (from e.g. protostellar jets) on the parent sub-clumps from which the stars form. With this assumption, we find an integrated star-formation efficiency (after all mass either turns into stars or is unbound) of $\sim 5-10\%$ for GMCs of all sizes. Interestingly, this is almost completely independent of the EOS we assume (either constant $\Sigma_{\rm crit}$ or $\Sigma_{\rm crit}\propto T^{2}$), as long as it terminates the fragmentation cascade at roughly the same point. Of course, if we assume this gas remains bound to the total system, so it is simply recycled back to the \myquote{top level} of the original fragmentation hierarchy until it is consumed (which obviously corresponds to a no-feedback case), then we trivially predict that eventually all gas turns into stars. Of course, the effects of realistic feedback are much more complex than these simplistic assumptions, and we could adopt arbitrarily complex models (for example, evolving each protostar and tracking explicitly location-dependent photo-ionization feedback, which we then use to explicitly calculate whether gas is unbound from the system). We note this result simply to demonstrate the utility of these semi-analytic models for rapidly exploring different assumptions regarding the effects of feedback.

\section{Conclusions}\label{sec:conclusions}

The aim of this paper is to provide a general framework for the modeling of star formation through turbulent fragmentation from the scale of GMCs to the scale of stars in order to quickly test the effects of different assumptions and new physics. Such a tool could allow theorists to explore different models and parameters before committing significant resources towards a detailed numerical simulation. We propose a semi analytical extension of the model of \cite{GuszejnovIMF} (\CMFIMF) that we believe is detailed enough to capture the physics essential for modeling the formation of stars without being too demanding numerically. Just like the analytical excursion set models it does not simulate turbulence directly, instead it assumes that the density follows a locally random field distribution whose parameters evolve in time so that virial equilibrium is satisfied. This is an assumption about turbulent collapse that needs to be tested in future work. The density field is directly resolved on a grid which preserves spatial and time information allowing the implementation of more detailed physics (e.g. proper checking for self gravitation, time dependent cloud collapse) and the analysis of the spatial structure. This is not possible in the excursion set formalism which describes the density field around a random Lagrangian point. This also means that unlike the analytical models not 100\% of the mass ends up in protostars.

The presented form of the model contains only the minimally required physics (turbulence, self gravity, some equation of state). It is however possible to integrate more sophisticated models to provide a more accurate description of these processes. Also, since the output of our model contain the time dependent evolution of the CMF and the PSMF, one can easily apply corrections during post processing to account for effects like protostellar fragmentation or supernova feedback (stop the evolution when enough SNe exploded).

By applying this framework to modeling the collapse of giant molecular clouds, we found that even the basic model qualitatively reproduces the observed core mass function. The CMF evolution has little dependence on the mass of the parent GMC mass.

Another result of the simulation is the mass distribution of all bound structures in the cloud. This appears to have the same shape as the CMF with a shallower slope of roughly $M^{-0.3}$ at the massive end. These clearly show the hierarchy of bound structures.

One of the main results of our basic model is the protostellar system mass function (PSMF) which is obtained by following the collapse of an ensemble of GMCs following a GMC mass function determined by \cite{core_IMF}. As in \CMFIMF we found that the PSMF is qualitatively very similar to the observed IMF: it exhibits a close to Salpeter slope almost independent of the initial conditions, while the turnover mass is mainly set by the equation of state and the initial temperature.

Due to the minimalistic nature of the model we managed to pinpoint the physical quantities influencing the different features of the PSMF and thus the IMF. We found that the Salpeter slope at the high mass end is a clear consequence of turbulence (as shown before in \CMFIMF) where the inclusion of extra physics only causes slight deviation from the pure power law behavior. Furthermore we found that in a medium with a stiff equation of state the actual turnover point in leading order is set by the local temperature ($M_{\rm crit}\propto T^2/\Sigma_{\rm crit}$).

We found that if we assume a $\gamma(\Sigma)$ equation of state then the PSMF for protostars of the same age changes as the parent cloud collapses: the turnover mass increases with time. This can be explained by the increase of $M_{\rm crit}$. This leads to a quadratic dependence of the turnover mass on the initial temperature which is inconsistent with the observed universality of the IMF. This means that it is not possible to derive a universal IMF with an equation of state that has no temperature dependence. One way to 'fix' the model is by implementing the feedback from protostars. Using the assumptions of \cite{Krumholz_stellar_mass_origin} in leading order the heating from the protostars cancel the aforementioned quadratic scaling (due to $\Sigma_{\rm crit}\propto T^2$), leading to a close to universal turnover mass.

\acknowledgments

We thank Ralf Klessen and Mark Krumholz for their insights and inspirational conversations throughout the development of this work.\\
Support for PFH and DG was provided by an Alfred P. Sloan Research Fellowship, NASA ATP Grant NNX14AH35G, and NSF Collaborative Research Grant \#1411920 and CAREER grant \#1455342. Numerical calculations were run on the Caltech computer cluster ``Zwicky'' (NSF MRI award \#PHY-0960291) and allocation TG-AST130039 granted by the Extreme Science and Engineering Discovery Environment (XSEDE) supported by the NSF. 

\bibliographystyle{mnras}
\bibliography{bibliography}

\begin{thebibliography}{}
\makeatletter
\relax
\def\mn@urlcharsother{\let\do\@makeother \do\$\do\&\do\#\do\^\do\_\do\%\do\~}
\def\mn@doi{\begingroup\mn@urlcharsother \@ifnextchar [ {\mn@doi@}
  {\mn@doi@[]}}
\def\mn@doi@[#1]#2{\def\@tempa{#1}\ifx\@tempa\@empty \href
  {http://dx.doi.org/#2} {doi:#2}\else \href {http://dx.doi.org/#2} {#1}\fi
  \endgroup}
\def\mn@eprint#1#2{\mn@eprint@#1:#2::\@nil}
\def\mn@eprint@arXiv#1{\href {http://arxiv.org/abs/#1} {{\tt arXiv:#1}}}
\def\mn@eprint@dblp#1{\href {http://dblp.uni-trier.de/rec/bibtex/#1.xml}
  {dblp:#1}}
\def\mn@eprint@#1:#2:#3:#4\@nil{\def\@tempa {#1}\def\@tempb {#2}\def\@tempc
  {#3}\ifx \@tempc \@empty \let \@tempc \@tempb \let \@tempb \@tempa \fi \ifx
  \@tempb \@empty \def\@tempb {arXiv}\fi \@ifundefined
  {mn@eprint@\@tempb}{\@tempb:\@tempc}{\expandafter \expandafter \csname
  mn@eprint@\@tempb\endcsname \expandafter{\@tempc}}}

\bibitem[\protect\citeauthoryear{{Arce}, {Shepherd}, {Gueth}, {Lee},
  {Bachiller}, {Rosen}  \& {Beuther}}{{Arce}
  et~al.}{2007}]{Arce_outflow_review}
{Arce} H.~G.,  {Shepherd} D.,  {Gueth} F.,  {Lee} C.-F.,  {Bachiller} R.,
  {Rosen} A.,   {Beuther} H.,  2007, Protostars and Planets V, \href
  {http://adsabs.harvard.edu/abs/2007prpl.conf..245A} {pp 245--260}

\bibitem[\protect\citeauthoryear{{Bate}}{{Bate}}{2009}]{Bate_2009_rad_importance}
{Bate} M.~R.,  2009, \mn@doi [\mnras] {10.1111/j.1365-2966.2008.14165.x}, \href
  {http://adsabs.harvard.edu/abs/2009MNRAS.392.1363B} {392, 1363}

\bibitem[\protect\citeauthoryear{{Bolatto}, {Leroy}, {Rosolowsky}, {Walter}  \&
  {Blitz}}{{Bolatto} et~al.}{2008}]{Bolatto_2008}
{Bolatto} A.~D.,  {Leroy} A.~K.,  {Rosolowsky} E.,  {Walter} F.,   {Blitz} L.,
  2008, \mn@doi [\apj] {10.1086/591513}, \href
  {http://adsabs.harvard.edu/abs/2008ApJ...686..948B} {686, 948}

\bibitem[\protect\citeauthoryear{{Chabrier}}{{Chabrier}}{2005}]{Chabrier_IMF}
{Chabrier} G.,  2005, in {Corbelli} E.,  {Palla} F.,   {Zinnecker} H.,  eds,
  Astrophysics and Space Science Library Vol. 327, The Initial Mass Function 50
  Years Later. p.~41

\bibitem[\protect\citeauthoryear{{Dale}, {Ercolano}  \& {Bonnell}}{{Dale}
  et~al.}{2012}]{Dale_ion_feedback}
{Dale} J.~E.,  {Ercolano} B.,   {Bonnell} I.~A.,  2012, \mn@doi [\mnras]
  {10.1111/j.1365-2966.2012.21205.x}, \href
  {http://adsabs.harvard.edu/abs/2012MNRAS.424..377D} {424, 377}

\bibitem[\protect\citeauthoryear{{Federrath}}{{Federrath}}{2013}]{Federrath_slope}
{Federrath} C.,  2013, \mn@doi [\mnras] {10.1093/mnras/stt1644}, \href
  {http://adsabs.harvard.edu/abs/2013MNRAS.436.1245F} {436, 1245}

\bibitem[\protect\citeauthoryear{{Federrath} \& {Klessen}}{{Federrath} \&
  {Klessen}}{2013}]{Federrath_density_distrib}
{Federrath} C.,  {Klessen} R.~S.,  2013, \mn@doi [\apj]
  {10.1088/0004-637X/763/1/51}, \href
  {http://adsabs.harvard.edu/abs/2013ApJ...763...51F} {763, 51}

\bibitem[\protect\citeauthoryear{{Federrath}, {Klessen}  \&
  {Schmidt}}{{Federrath} et~al.}{2008}]{Federrath_turbulence_compressive_PDF}
{Federrath} C.,  {Klessen} R.~S.,   {Schmidt} W.,  2008, \mn@doi [\apjl]
  {10.1086/595280}, \href {http://adsabs.harvard.edu/abs/2008ApJ...688L..79F}
  {688, L79}

\bibitem[\protect\citeauthoryear{{Geen}, {Rosdahl}, {Blaizot}, {Devriendt}  \&
  {Slyz}}{{Geen} et~al.}{2015}]{Geen_massive_feedback}
{Geen} S.,  {Rosdahl} J.,  {Blaizot} J.,  {Devriendt} J.,   {Slyz} A.,  2015,
  \mn@doi [\mnras] {10.1093/mnras/stv251}, \href
  {http://adsabs.harvard.edu/abs/2015MNRAS.448.3248G} {448, 3248}

\bibitem[\protect\citeauthoryear{{Glover} \& {Mac Low}}{{Glover} \& {Mac
  Low}}{2007}]{Glover_EQS_lowgamma_ref}
{Glover} S.~C.~O.,  {Mac Low} M.-M.,  2007, \mn@doi [\apjs] {10.1086/512238},
  \href {http://adsabs.harvard.edu/abs/2007ApJS..169..239G} {169, 239}

\bibitem[\protect\citeauthoryear{{Goodwin}, {Whitworth}  \&
  {Ward-Thompson}}{{Goodwin} et~al.}{2004}]{Goodwin04a}
{Goodwin} S.~P.,  {Whitworth} A.~P.,   {Ward-Thompson} D.,  2004, \aap, 414,
  633

\bibitem[\protect\citeauthoryear{Guszejnov \& Hopkins}{Guszejnov \&
  Hopkins}{2015}]{GuszejnovIMF}
Guszejnov D.,  Hopkins P.~F.,  2015, \mn@doi [\mnras] {10.1093/mnras/stv872},
  450, 4137

\bibitem[\protect\citeauthoryear{{Guszejnov}, {Krumholz}  \&
  {Hopkins}}{{Guszejnov} et~al.}{2015}]{guszejnov_feedback_necessity}
{Guszejnov} D.,  {Krumholz} M.~R.,   {Hopkins} P.~F.,  2015, preprint, \href
  {http://adsabs.harvard.edu/abs/2015arXiv151005040G} {} (\mn@eprint {arXiv}
  {1510.05040})

\bibitem[\protect\citeauthoryear{{Hanawa} \& {Matsumoto}}{{Hanawa} \&
  {Matsumoto}}{1999}]{Hanawa_bar_perturbation}
{Hanawa} T.,  {Matsumoto} T.,  1999, \mn@doi [\apj] {10.1086/307564}, \href
  {http://adsabs.harvard.edu/abs/1999ApJ...521..703H} {521, 703}

\bibitem[\protect\citeauthoryear{{Hennebelle}}{{Hennebelle}}{2012}]{Hennebelle2012}
{Hennebelle} P.,  2012, \mn@doi [\aap] {10.1051/0004-6361/201219440}, \href
  {http://adsabs.harvard.edu/abs/2012A%26A...545A.147H} {545, A147}

\bibitem[\protect\citeauthoryear{{Hennebelle} \& {Chabrier}}{{Hennebelle} \&
  {Chabrier}}{2008}]{HC08}
{Hennebelle} P.,  {Chabrier} G.,  2008, \mn@doi [\apj] {10.1086/589916}, \href
  {http://adsabs.harvard.edu/abs/2008ApJ...684..395H} {684, 395}

\bibitem[\protect\citeauthoryear{{Hopkins}}{{Hopkins}}{2012a}]{excursion_set_ism}
{Hopkins} P.~F.,  2012a, \mn@doi [\mnras] {10.1111/j.1365-2966.2012.20730.x},
  \href {http://adsabs.harvard.edu/abs/2012MNRAS.423.2016H} {423, 2016}

\bibitem[\protect\citeauthoryear{{Hopkins}}{{Hopkins}}{2012b}]{core_IMF}
{Hopkins} P.~F.,  2012b, \mn@doi [\mnras] {10.1111/j.1365-2966.2012.20731.x},
  \href {http://adsabs.harvard.edu/abs/2012MNRAS.423.2037H} {423, 2037}

\bibitem[\protect\citeauthoryear{{Hopkins}}{{Hopkins}}{2013a}]{general_turbulent_fragment}
{Hopkins} P.~F.,  2013a, \mn@doi [\mnras] {10.1093/mnras/sts704}, \href
  {http://adsabs.harvard.edu/abs/2013MNRAS.430.1653H} {430, 1653}

\bibitem[\protect\citeauthoryear{{Hopkins}}{{Hopkins}}{2013b}]{Hopkins_isothermal_turb}
{Hopkins} P.~F.,  2013b, \mn@doi [\mnras] {10.1093/mnras/stt010}, \href
  {http://adsabs.harvard.edu/abs/2013MNRAS.430.1880H} {430, 1880}

\bibitem[\protect\citeauthoryear{{Iffrig} \& {Hennebelle}}{{Iffrig} \&
  {Hennebelle}}{2015}]{Iffrig_SN_feedback}
{Iffrig} O.,  {Hennebelle} P.,  2015, \mn@doi [\aap]
  {10.1051/0004-6361/201424556}, \href
  {http://adsabs.harvard.edu/abs/2015A%26A...576A..95I} {576, A95}

\bibitem[\protect\citeauthoryear{{Jappsen}, {Klessen}, {Larson}, {Li}  \& {Mac
  Low}}{{Jappsen} et~al.}{2005}]{Jappsen_EQS_ref}
{Jappsen} A.-K.,  {Klessen} R.~S.,  {Larson} R.~B.,  {Li} Y.,   {Mac Low}
  M.-M.,  2005, \mn@doi [\aap] {10.1051/0004-6361:20042178}, \href
  {http://adsabs.harvard.edu/abs/2005A%26A...435..611J} {435, 611}

\bibitem[\protect\citeauthoryear{{Kroupa}}{{Kroupa}}{2002}]{Kroupa_IMF}
{Kroupa} P.,  2002, \mn@doi [Science] {10.1126/science.1067524}, \href
  {http://adsabs.harvard.edu/abs/2002Sci...295...82K} {295, 82}

\bibitem[\protect\citeauthoryear{{Krumholz}}{{Krumholz}}{2011}]{Krumholz_stellar_mass_origin}
{Krumholz} M.~R.,  2011, \mn@doi [\apj] {10.1088/0004-637X/743/2/110}, \href
  {http://adsabs.harvard.edu/abs/2011ApJ...743..110K} {743, 110}

\bibitem[\protect\citeauthoryear{{Krumholz}}{{Krumholz}}{2014}]{SF_big_problems}
{Krumholz} M.~R.,  2014, \mn@doi [\physrep] {10.1016/j.physrep.2014.02.001},
  \href {http://adsabs.harvard.edu/abs/2014PhR...539...49K} {539, 49}

\bibitem[\protect\citeauthoryear{{Lada} \& {Lada}}{{Lada} \&
  {Lada}}{2003}]{clustering_Lada}
{Lada} C.~J.,  {Lada} E.~A.,  2003, \mn@doi [\araa]
  {10.1146/annurev.astro.41.011802.094844}, \href
  {http://adsabs.harvard.edu/abs/2003ARA%26A..41...57L} {41, 57}

\bibitem[\protect\citeauthoryear{{Larson}}{{Larson}}{1981}]{Larson_law}
{Larson} R.~B.,  1981, \mnras, \href
  {http://adsabs.harvard.edu/abs/1981MNRAS.194..809L} {194, 809}

\bibitem[\protect\citeauthoryear{{Masunaga} \& {Inutsuka}}{{Masunaga} \&
  {Inutsuka}}{2000}]{Masunaga_EQS_highgamma_ref}
{Masunaga} H.,  {Inutsuka} S.-i.,  2000, \mn@doi [\apj] {10.1086/308439}, \href
  {http://adsabs.harvard.edu/abs/2000ApJ...531..350M} {531, 350}

\bibitem[\protect\citeauthoryear{{McKee} \& {Offner}}{{McKee} \&
  {Offner}}{2010}]{Offner_protostar_MF}
{McKee} C.~F.,  {Offner} S.~S.~R.,  2010, \mn@doi [\apj]
  {10.1088/0004-637X/716/1/167}, \href
  {http://adsabs.harvard.edu/abs/2010ApJ...716..167M} {716, 167}

\bibitem[\protect\citeauthoryear{{McKee} \& {Ostriker}}{{McKee} \&
  {Ostriker}}{2007a}]{McKee_star_formation}
{McKee} C.~F.,  {Ostriker} E.~C.,  2007a, \mn@doi [\araa]
  {10.1146/annurev.astro.45.051806.110602}, \href
  {http://adsabs.harvard.edu/abs/2007ARA%26A..45..565M} {45, 565}

\bibitem[\protect\citeauthoryear{{McKee} \& {Ostriker}}{{McKee} \&
  {Ostriker}}{2007b}]{McKee_SF_theory}
{McKee} C.~F.,  {Ostriker} E.~C.,  2007b, \mn@doi [\araa]
  {10.1146/annurev.astro.45.051806.110602}, \href
  {http://adsabs.harvard.edu/abs/2007ARA%26A..45..565M} {45, 565}

\bibitem[\protect\citeauthoryear{{Murray}}{{Murray}}{1973}]{Murray_supersonic_Burgers_analytic}
{Murray} J.~D.,  1973, \mn@doi [Journal of Fluid Mechanics]
  {10.1017/S0022112073001564}, \href
  {http://adsabs.harvard.edu/abs/1973JFM....59..263M} {59, 263}

\bibitem[\protect\citeauthoryear{Murray \& Chang}{Murray \&
  Chang}{2015}]{Murray_star_formation}
Murray N.,  Chang P.,  2015, \apj, 804, 44

\bibitem[\protect\citeauthoryear{{Murray}, {Chang}, {Murray}  \&
  {Pittman}}{{Murray} et~al.}{2015}]{Murray_2015_turb_sim}
{Murray} D.~W.,  {Chang} P.,  {Murray} N.~W.,   {Pittman} J.,  2015, preprint,
  \href {http://adsabs.harvard.edu/abs/2015arXiv150905910M} {} (\mn@eprint
  {arXiv} {1509.05910})

\bibitem[\protect\citeauthoryear{{Nakano} \& {Nakamura}}{{Nakano} \&
  {Nakamura}}{1978}]{Nakano_magnetic_fields}
{Nakano} T.,  {Nakamura} T.,  1978, \pasj, \href
  {http://adsabs.harvard.edu/abs/1978PASJ...30..671N} {30, 671}

\bibitem[\protect\citeauthoryear{{Ntormousi} \& {Hennebelle}}{{Ntormousi} \&
  {Hennebelle}}{2015}]{Ntormousi_core_shell_instability}
{Ntormousi} E.,  {Hennebelle} P.,  2015, \mn@doi [\aap]
  {10.1051/0004-6361/201424705}, \href
  {http://adsabs.harvard.edu/abs/2015A%26A...574A.130N} {574, A130}

\bibitem[\protect\citeauthoryear{{Offner}, {Clark}, {Hennebelle}, {Bastian},
  {Bate}, {Hopkins}, {Moraux}  \& {Whitworth}}{{Offner}
  et~al.}{2014}]{IMF_universality}
{Offner} S.~S.~R.,  {Clark} P.~C.,  {Hennebelle} P.,  {Bastian} N.,  {Bate}
  M.~R.,  {Hopkins} P.~F.,  {Moraux} E.,   {Whitworth} A.~P.,  2014, \mn@doi
  [Protostars and Planets VI] {10.2458/azu_uapress_9780816531240-ch003}, \href
  {http://adsabs.harvard.edu/abs/2014prpl.conf...53O} {pp 53--75}

\bibitem[\protect\citeauthoryear{{Padoan} \& {Nordlund}}{{Padoan} \&
  {Nordlund}}{2002}]{Padoan_Nordlund_2002_IMF}
{Padoan} P.,  {Nordlund} {\AA}.,  2002, \mn@doi [\apj] {10.1086/341790}, \href
  {http://adsabs.harvard.edu/abs/2002ApJ...576..870P} {576, 870}

\bibitem[\protect\citeauthoryear{{Padoan}, {Nordlund}  \& {Jones}}{{Padoan}
  et~al.}{1997}]{Padoan_theory}
{Padoan} P.,  {Nordlund} A.,   {Jones} B.~J.~T.,  1997, \mnras, \href
  {http://adsabs.harvard.edu/abs/1997MNRAS.288..145P} {288, 145}

\bibitem[\protect\citeauthoryear{{Press} \& {Schechter}}{{Press} \&
  {Schechter}}{1974}]{PressSchechter}
{Press} W.~H.,  {Schechter} P.,  1974, \mn@doi [\apj] {10.1086/152650}, \href
  {http://adsabs.harvard.edu/abs/1974ApJ...187..425P} {187, 425}

\bibitem[\protect\citeauthoryear{{Robertson} \& {Goldreich}}{{Robertson} \&
  {Goldreich}}{2012}]{Brant_turb_pumping}
{Robertson} B.,  {Goldreich} P.,  2012, \mn@doi [\apjl]
  {10.1088/2041-8205/750/2/L31}, \href
  {http://adsabs.harvard.edu/abs/2012ApJ...750L..31R} {750, L31}

\bibitem[\protect\citeauthoryear{{Rosolowsky}}{{Rosolowsky}}{2005}]{Rosolowsky_2005_GMC}
{Rosolowsky} E.,  2005, \mn@doi [\pasp] {10.1086/497582}, \href
  {http://adsabs.harvard.edu/abs/2005PASP..117.1403R} {117, 1403}

\bibitem[\protect\citeauthoryear{{Rosolowsky}, {Pineda}, {Kauffmann}  \&
  {Goodman}}{{Rosolowsky} et~al.}{2008}]{Rosolowsky2008_dendogram}
{Rosolowsky} E.~W.,  {Pineda} J.~E.,  {Kauffmann} J.,   {Goodman} A.~A.,  2008,
  \mn@doi [\apj] {10.1086/587685}, \href
  {http://adsabs.harvard.edu/abs/2008ApJ...679.1338R} {679, 1338}

\bibitem[\protect\citeauthoryear{{Sadavoy} et~al.,}{{Sadavoy}
  et~al.}{2010}]{Sadavoy_observed_CMF}
{Sadavoy} S.~I.,  et~al., 2010, \mn@doi [\apj] {10.1088/0004-637X/710/2/1247},
  \href {http://adsabs.harvard.edu/abs/2010ApJ...710.1247S} {710, 1247}

\bibitem[\protect\citeauthoryear{{Schmidt}, {Federrath}, {Hupp}, {Kern}  \&
  {Niemeyer}}{{Schmidt} et~al.}{2009}]{Schmidt_supersonic_sim}
{Schmidt} W.,  {Federrath} C.,  {Hupp} M.,  {Kern} S.,   {Niemeyer} J.~C.,
  2009, \mn@doi [\aap] {10.1051/0004-6361:200809967}, \href
  {http://adsabs.harvard.edu/abs/2009A%26A...494..127S} {494, 127}

\bibitem[\protect\citeauthoryear{{Walch}, {Whitworth}  \& {Girichidis}}{{Walch}
  et~al.}{2012a}]{Walch12a}
{Walch} S.,  {Whitworth} A.~P.,   {Girichidis} P.,  2012a, \mnras, 419, 760

\bibitem[\protect\citeauthoryear{{Walch}, {Whitworth}, {Bisbas}, {W{\"u}nsch}
  \& {Hubber}}{{Walch} et~al.}{2012b}]{Walch_ion_feedback}
{Walch} S.~K.,  {Whitworth} A.~P.,  {Bisbas} T.,  {W{\"u}nsch} R.,   {Hubber}
  D.,  2012b, \mn@doi [\mnras] {10.1111/j.1365-2966.2012.21767.x}, \href
  {http://adsabs.harvard.edu/abs/2012MNRAS.427..625W} {427, 625}

\bibitem[\protect\citeauthoryear{{Zamora-Avil{\'e}s}, {V{\'a}zquez-Semadeni}
  \& {Col{\'{\i}}n}}{{Zamora-Avil{\'e}s} et~al.}{2012}]{Vazquez_SF_model_2012}
{Zamora-Avil{\'e}s} M.,  {V{\'a}zquez-Semadeni} E.,   {Col{\'{\i}}n} P.,  2012,
  \mn@doi [\apj] {10.1088/0004-637X/751/1/77}, \href
  {http://adsabs.harvard.edu/abs/2012ApJ...751...77Z} {751, 77}

\makeatother
\end{thebibliography}

\appendix

\section{Basic Simulation Algorithm}\label{sec:algorithm}

\begin{figure*}
\begin{tikzpicture}[ node distance=3cm]

\node (initialize) [startstop] {{\bf 1}\\Start cloud evolution with $\rho(\mathbf{x})$, $T(\mathbf{x})$, $R$ and $\mach_{\rm edge}$ as input. Initialize turbulence. Prepare $\delta\left(\mathbf{k},0\right)$ using FFT.};

\node (tstep) [process, below of=initialize] {{\bf 2}\\Evolve global parameters and $\delta\left(\mathbf{k},t\right)$ one time step. Calculate $\rho$ and $T$ fields at $t+{\rm d}t$.};

\node (dec_fragment) [decision,below of=tstep , yshift=-1 cm] {{\bf 3}\\Is there a fragmenting subregion?};

\node (dec_selfgrav) [decision,below of=dec_fragment , yshift=-1 cm] {{\bf 5 (c)}\\Is the cloud still self-gravitating?};

\node (dec_time)  [decision,below of=dec_selfgrav , yshift=-1 cm] {{\bf 5 (a,b)}\\Collapse criteria met?};

\node (recursion) [process, right of=tstep, xshift=2.5cm] {{\bf 4}\\Recursively follow the subregion using current state as initial conditions.};

\node (return_data) [startstop, right of=dec_selfgrav, xshift=2.5cm] {{\bf 5(c)}\\ Continue evolving for one dynamical time. If no fragments have formed throw away the unbound material.};

\node (collapse) [startstop, right of=dec_time, xshift=2.5cm] {{\bf 6}\\ The cloud collapsed to a protostar. Write data, return to parent.};

\draw [arrow] (initialize) -- (tstep);
\draw [arrow] (tstep) -- (dec_fragment);

\draw [arrow] (dec_fragment) -- node[anchor=east] {no} (dec_selfgrav);
\draw [arrow] (dec_fragment) -| node[anchor=west] {yes} (recursion);

\draw [arrow] (dec_selfgrav) -- node[anchor=north] {no} (return_data);
\draw [arrow] (dec_selfgrav) -- node[anchor=east] {yes} (dec_time);

\draw [arrow] (dec_time) -- node[anchor=north] {yes} (collapse);

\draw [arrow] (recursion) |-  (initialize);


\draw[->, thick] (dec_time) -- node[anchor=north] {no} ($(dec_time.west)+(-2,0)$) |- (tstep) ;

\end{tikzpicture}
\caption{Basic algorithm of fragmentation code. The bold numbers in each box show which step from Appendix \ref{sec:algorithm} they represent. See Appendix \ref{sec:algorithm} for more detailed description.}\label{flowchart}
\end{figure*}
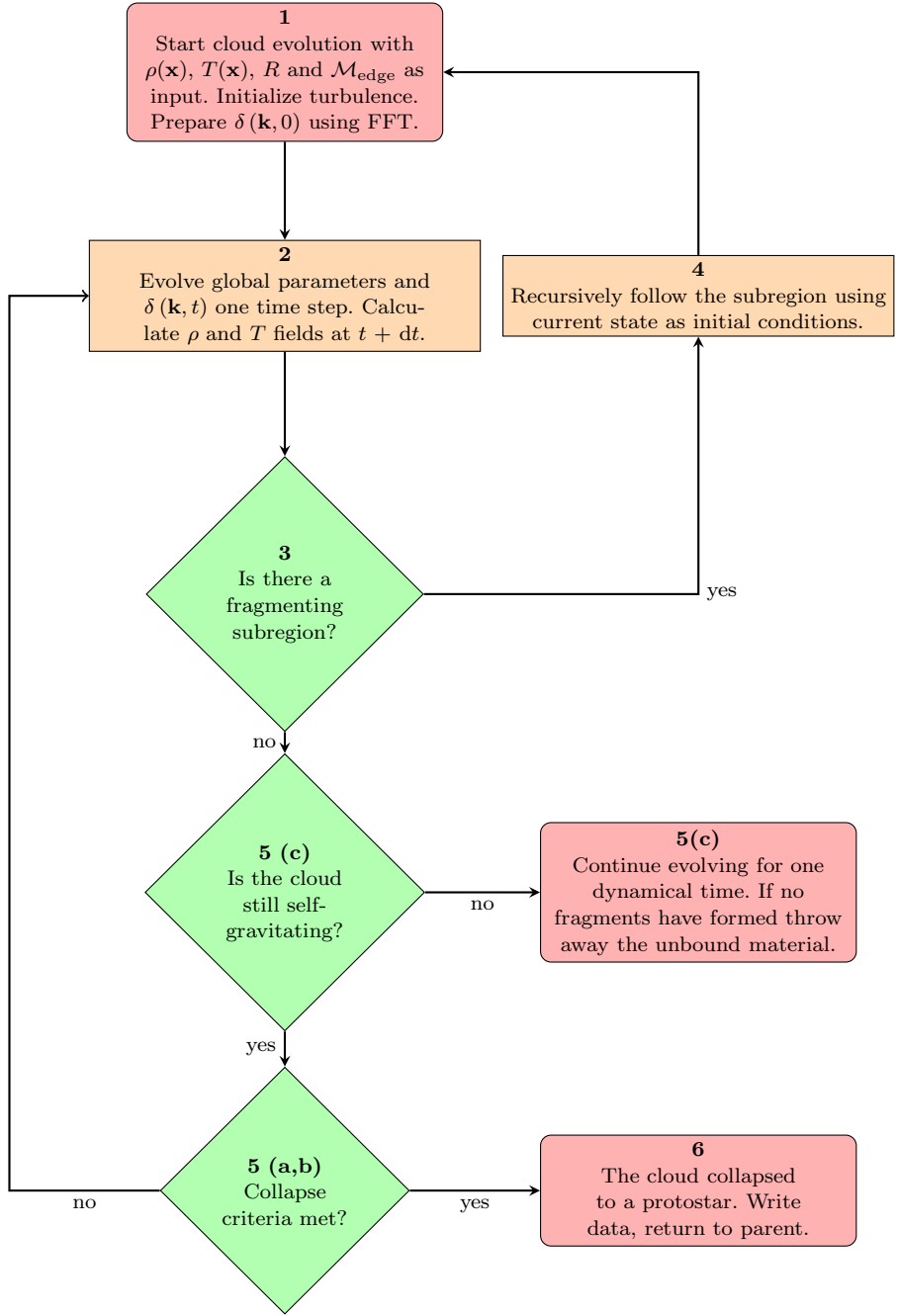

In this appendix we detail step-by-step how the basic version of the simulation works (see flowchart of Fig. \ref{flowchart}), but note that it can be greatly expanded with new physics, as long as the fundamental assumption (locally random density modes) is kept. 
\begin{enumerate}[label=(\arabic*)]
	\item We begin with a GMC sized cloud whose initial parameters (mass, radius, temperature, density, edge Mach number, sound speed etc.) are derived from its mass ($M$), the sonic mass ($M_{\rm sonic}$) and length ($R_{\rm sonic}$), using the mass-size relation of Eq. \ref{eq:mass_size} and linewidth-size relation of Eq. \ref{eq:mach_scaling}. These are all initialized on a 3D spatial grid, of resolution $N\times N\times N$ chosen such that the final statistics converge (we found this happens at $N\geq 16$). The density field is initialized assuming that it is lognormal (variance set according to Eq. \ref{eq:S_per_mode}) using the full density power spectrum model (transforming to Fourier space and back), while the temperature field follows the density according to the desired equation of state.
	
	\item \label{step2} We take timestep $\Delta t$ ($\Delta t\ll t_{\rm dyn}$ and $\Delta t\ll t_{\rm cross}(d)$, where $d=2R/N$ is the spatial resolution of the grid). This means:
	\begin{enumerate}[label=(\alph*)]
		\item Global contraction of the cloud (all scales shrink, density uniformly increases) according to Eq. \ref{eq:scale_evol}.
		\item The density perturbation power spectrum $\delta(\kvect)$ is updated following Eq. \ref{eq:delta_evol}, which assumes density mode statistics obey a local ``random walk'' in phase space. The actual density field is calculated by Fourier transforming to real space and normalizing the field with the cloud mass (this way mass is conserved).
		\item The temperature field is updated according to new densities and the chosen EOS (see Eq. \ref{eq:T_evol}).
		\item The cloud scale Mach number is updated according to our assumption that the virial parameter is constant during collapse.
	\end{enumerate}
	
	\item We now check whether any self-gravitating substructures have formed by using a Monte Carlo method that involves placing spheres of all possible sizes at random positions and testing them using the collapse criterion of Eq. \ref{eq:collapse_threshold_T}.
	
	\item \label{step4} If such a region is found it is ``removed'' temporarily and expanded into its own grid. This new grid will have a higher spatial resolution than its parent, thus density modes on the newly available small scales need to be initialized (larger modes are inherited from previous grid). We then repeat steps \ref{step2}-\ref{step4} on this new grid. This means that during the evolution of its fragments the parent cloud is ``frozen'' in time. This is motivated by the fact that the dynamical time of fragments is smaller as $t_{\rm dyn}\propto 1/\sqrt{\rho}$, so they evolve ``fast'' compared to their parents. Note that all clouds keep track of physical time, so it is possible to properly date the formation times of protostars and clouds. 
	
	\item The time evolution of each cloud/grid continues until:
\begin{enumerate}[label=(\alph*)]
	\item The cloud reaches the protostellar size scale ($R<R_{\rm min}$), below which it is assumed to have formed a protostar.
	\item The cloud is still self gravitating after a number of dynamical times ($t>t_{\rm max}$)\footnote{This can happen if $\gamma>1$, as $\tilde{r}$ in Eq. \ref{eq:scale_evol} does not reach zero  in a finite amount of time.}. After this limit is reached the cloud is assumed to have cooled and collapsed through other means. Essentially, this represents non-fragmenting cores. 
	\item The cloud stops being self-gravitating. This can happen if a cloud loses enough of its mass that it becomes unbound. Since virial equilibrium is enforced this means no turbulence, which means no more fragmentation. In the model presented above these clouds are \emph{not} forming stars or contributing to the mass of the protostars forming from their fragments, instead this material is ``thrown away'' (this represents ``feedback'' in some sense, see Sec. \ref{sec:model_diff}). Note that it is possible within the framework to return this unbound material to the parent GMC where it may form stars, but for simplicity in the presented model we chose not to do that.
\end{enumerate}

\item  Clouds that formed protostars are removed the properties of the protostars are cataloged. We then return to the parent cloud and continue its evolution from Step \ref{step4}.

\item This continues until 100\% of the original original mass of the cloud is either in protostars or unbound. The final output is the catalog of protostars. Note that it is also possible to get the CMF by exporting the properties of bound structures at a specified time. The whole process is repeated for large number of initial GMCs (with different random seeds) to gain adequate statistics.
\end{enumerate}

\end{document}